%% file: manuscript.tex
\documentclass[ reprint, amsmath,amssymb,aps, pra]{revtex4-2}
\usepackage{hyperref}
\usepackage{graphicx}% Include figure files
\usepackage{dcolumn}% Align table columns on decimal point
\usepackage{bm}% bold math
\usepackage{xcolor}
\usepackage{cases}
\usepackage{float}
\usepackage[overload]{empheq}

\newcommand{\imag}[1]{\text{Im}\{ #1 \}}
\newcommand{\real}[1]{\text{Re}\left\{ #1 \right\}}
\newcommand{\betamax}{\beta_{\text{max}}}
\newcommand{\Li}[1]{\text{Li}_2\left[ #1 \right]}
\newcommand{\bomega }{(\omega)}
\newcommand{\epsdrude}{\varepsilon_{\text{plasm}}}
\newcommand{\epsdl}{\varepsilon_{\text{polar}}}
\newcommand{\Omegaplasm}{\Omega_{\text{plasm}}}
\newcommand{\Omegapolar}{\Omega_{\text{polar}}}
\newcommand{\rads}{\,\text{rad/s}}
\newcommand{\hmax}{h_\text{max}}
\newcommand{\Bplasm}{B_\text{plasm}}
\newcommand{\Bpolar}{B_\text{polar}}
\newcommand{\Qopt}{Q_\text{opt}}
\newcommand{\Topt}{T_\text{opt}}
\newcommand{\bNF}{b_\text{NF}}
\newcommand{\bWien}{b_\text{Wien}}
\newcommand{\PiT}{\Pi\left( \frac{\Omega}{T} \right)}

\newcommand{\etal}{\textit{et al. }}

\newcommand{\Qth}{Q_\text{th}}
\newcommand{\Fplasm}{F_\text{plasm}}
\newcommand{\Fpolar}{F_\text{polar}}

\newcommand{\res}{\mathcal{R}\text{es}}

\begin{document}

\preprint{APS/123-QED}

\title{Tight bounds and the role of optical loss in polariton-mediated near-field heat transfer} 
\author{Mariano Pascale}
\email{mariano.pascale@icfo.eu}
\author{Georgia T. Papadakis}
\affiliation{%
ICFO-Institut de Ciencies Fotoniques, The Barcelona Institute of Science and Technology, Castelldefels (Barcelona) 08860, Spain
}%

\begin{abstract}
We introduce an analytical framework for near-field radiative heat transfer in bulk plasmonic and polar media. Considering material dispersion, we derive a closed form expression for the radiative thermal conductance, which disentangles the role of optical loss from other material dispersion characteristics, such as the spectral width of the Reststrahlen band in polar dielectrics, as well as from the temperature. We provide a universal condition for maximizing heat transfer that defines the optimal interplay between a material's optical loss and polariton resonance frequency, based on which we introduce tight bounds to near-field heat transfer. With this formalism, one can quantitatively evaluate all polaritonic materials in terms of their performance as near-field thermal emitters.
\end{abstract}
%\keywords{Suggested keywords}%Use showkeys class option if keyword                              %display desired
\maketitle
	
\section{Introduction} 

Radiative heat transfer between bodies separated by nanometric vacuum gaps can surpass the blackbody limit by several orders of magnitude \cite{rytov_theory_1953,polder_theory_1971,xu_heat_1994,kittel_near-field_2005,rousseau_radiative_2009,biehs_mesoscopic_2010,basu_review_2009}. This creates opportunities in a wide range of applications where thermal control is critical. Examples include high-efficiency energy conversion with thermophotovoltaic systems \cite{papadakis_broadening_2020,papadakis_thermodynamics_2021,molesky_high_2013,zhao_near-field_2018,francoeur_thermal_2011,desutter_external_2017},
contactless cooling \cite{kerschbaumer_contactless_2021,epstein_observation_1995}, thermal lithography \cite{pendry_radiative_1999,howell_thermal_2020,garcia_advanced_2014,hu_tip-based_2017}, thermally-assisted magnetic recording \cite{hamann_thermally_2004,ruigrok_disk_2000,kief_materials_2018}, and thermal logic circuitry \cite{otey_thermal_2010,fiorino_thermal_2018,ben-abdallah_near-field_2014,papadakis_deep-subwavelength_2021}.

Accessing the thermal near-field, where the separation distance between objects is smaller than the relevant thermal wavelength, typically on the order of few microns, yields ultra-high radiative heat transfer (on the order of tens of W/cm$^2$ \cite{hargreaves_anomalous_1969,domoto_experimental_1970,polder_theory_1971,mulet_enhanced_2002,rousseau_radiative_2009,papadakis_gate-tunable_2019}). In this range, radiative heat is optimally transferred by evanescent modes, such as surface plasmon polaritons (SPPs) and surface phonon polaritons (SPhPs) \cite{mulet_enhanced_2002,caldwell_low-loss_2015}, that tunnel across a nanometric vacuum gap. SPPs are resonantly excited on a plasmonic metal surface \cite{maier_plasmonics_2007}, e.g., Au or Ag, at frequencies ranging from near-infrared (IR) to ultraviolet \cite{kim_plasmonic_2013}. On the other hand, SPhPs are surface resonant modes occurring on polar dielectrics and semiconductors, e.g., hexagonal BN, SiC or GaAs, at frequencies ranging from mid-IR to $<10\,\mathrm{THz}$ \cite{caldwell_low-loss_2015}.

 Near-field radiative heat transfer (NFRHT) is rigorously described within the framework of fluctuational electrodynamics (FE) \cite{rytov_theory_1953,polder_theory_1971}. Computing the total NFRHT between bodies requires carrying out a spectral as well as a spatial integration of the exchanged thermal radiation over the entire frequency spectrum and for all relevant wavenumbers, respectively. This integration remains challenging in practice even in the simplest case of NFRHT between semi-infinite planar layers, for which numerical approaches are usually employed \cite{wang_parametric_2009,chen_mesh_2018}. {Most importantly, although optical loss and thermal fluctuations are fundamentally connected via the fluctuation-dissipation theorem \cite{rytov_theory_1953}, the dependence of NFRHT on the optical loss of materials remains unidentified.}
 
 { Upper bounds to the thermal emission spectrum per frequency (termed $\Phi(\omega)$ , Eq. (\ref{eq:Phi})) have been recently analytically estimated  \cite{pendry_radiative_1999,ben-abdallah_fundamental_2010,miller_shape-independent_2015, venkataram_fundamental_2020,molesky_fundamental_2020}. Furthermore, upper bounds to the integrated radiative thermal conductance (Eq. \ref{eq:HTC_def}) have been reported in \cite{zhang_all_2022,venkataram_fundamental_2020,ben-abdallah_fundamental_2010}.} In \cite{rousseau_asymptotic_2012,iizuka_analytical_2015}, analytical solutions to the total integrated heat transfer for polaritonic media were reported. Nonetheless, critical material parameters such as the polariton resonance frequency, the optical loss, as well as other dispersion characteristics, have remained mathematically intertwined. This prohibits a deep physical understanding of the role of each in NFRHT. 
 % To the authors knowledge, in no previous works have these parameters been disentangled to offer new insights into this otherwise well-studied problem in physics.

In this work, we provide a general and fully analytical framework for NFRHT mediated by surface polaritons in planar bulk systems.
Thus, we identify, the explicit dependence of NFRHT on material loss, and thereby derive a universal optimal loss condition that maximizes NFRHT. 
With respect to previous contributions \cite{venkataram_fundamental_2020,ben-abdallah_fundamental_2010}, by taking into account actual material dispersion, we introduce a temperature-independent tight upper bound to NFRHT. 
Furthermore, we carry out a quantitative classification of all polaritonic media in terms of NFRHT performance.  
Although the presented analytical framework models materials characterized by a single polariton resonance, we also show a semi-analytical extension able to describe multiple polaritonic resonances.

\section{Results and discussion} 
\subsection{Emission spectrum}
We describe NFRHT at a mean temperature, $T$, by evaluating the radiative thermal conductance per unit area \cite{song_near-field_2015}:
\begin{equation}\label{eq:HTC_def}
    h=\int_0^{\infty} \left[\frac{\partial}{\partial T}\theta(\omega,T)\right]\Phi\bomega d\omega,
\end{equation}
where $\theta(\omega,T)=\hbar\omega / \left[\exp{\left(\hbar\omega/k_B T\right)}-1\right]$ is mean energy per photon and $\Phi\bomega $ is the thermal emission spectrum, expressed in $\mathrm{m}^{-2}$. We consider a vacuum gap of size $d$ that separates two planar semi-infinite bodies exchanging heat. 
The bodies are made of nonmagnetic, isotropic, homogeneous material with relative dielectric permittivity $\varepsilon\bomega$  and susceptibility $\chi\bomega =\varepsilon\bomega -1$.
From fluctuational-electrodynamics \cite{rytov_theory_1953,polder_theory_1971,loomis_theory_1994,shchegrov_near-field_2000}, $\Phi\bomega $ is given by:
\begin{equation}\label{eq:Phi}
    \Phi\bomega  =\frac{1}{4\pi^2}\int_0^\infty [\xi_p (\omega,\beta)+\xi_s(\omega,\beta)] \beta d\beta,
\end{equation}
where $\xi_{p,s}(\omega,\beta)$ is the probability for a photon at frequency $\omega$ and in-plane wavenumber $\beta$ to tunnel accross the gap. The subscripts $p$ and $s$ denote polarization, corresponding to TM (transverse magnetic) and TE (transverse electric), respectively.

For gap sizes smaller than the thermal wavelength, $\lambda_\mathrm{T}=\bWien/T$, where $\bWien =2989\,\mu m \, K$ \cite{planck_zur_1901}, thermally excited SPPs and SPhPs dominate NFRHT in plasmonic and polar materials, respectively. Since these can only be excited in $p$-polarization \cite{maier_plasmonics_2007}, the emission spectrum, $\Phi$, can be approximated as the contribution from $p$-polarization alone. At sufficiently small vacuum gaps, the dispersion of surface polaritons approaches the quasistatic limit, for which $\beta\gg k_0$ \cite{pendry_radiative_1999}, where $k_0=\omega/c$ is the free-space wavenumber. 
In this limit, as shown in Eq. \eqref{Seq:beta_xi1} of Appendix \ref{appendixA}, the maximum in-plane wavenumber, $\betamax$, that satisfies the perfect photon tunneling condition, i.e., $\xi_p=1$, occurs near the surface polariton resonance frequency, $\Omega$, at which $\real{r_p}=0$, where $\displaystyle    r_p\bomega  = \frac{\chi\bomega }{\chi\bomega  +2}$ is the Fresnel coefficient in the quasistatic regime, or:
\begin{equation}\label{eq:resonance_freq}
 \real{\frac{2}{\chi (\Omega)}}=-1.
\end{equation}
For low-loss materials, i.e., for $\imag{\chi(\Omega)}\ll 1$, Eq.~\eqref{eq:resonance_freq} reduces to the more common expression $\real{\chi(\Omega)}=-2$ \cite{maier_plasmonics_2007}.

To obtain the emission spectrum, we carry out the integration of Eq.~\eqref{eq:Phi}, over all available wavenumbers, $\beta$. Upon assuming $\beta\gg k_0$, this integration yields: 
\begin{equation}\label{eq:Phi_analytical}
    \Phi\bomega = \frac{1}{8\pi^2d^2}\frac{\imag{r_p\bomega }}{\real{r_p\bomega }}\text{Im}\left\{\Li{r_p^2\bomega }\right\},
\end{equation}
where $\text{Li}_2$ is the dilogarithm or Spence's function \cite{lewin_dilogarithms_1958}, given in Eq. \eqref{Seq:dilog} of Appendix \ref{appendixA}. This expression agrees with Rousseau \etal \cite{rousseau_asymptotic_2012}. Its derivation, along with the more general expression for heat exchange between dissimilar materials, can be found in Appendix \ref{appendixA}, where we also showcase the validity of Eq.~\eqref{eq:Phi_analytical}. Eq.~\eqref{eq:Phi_analytical} directly computes the thermal emission spectrum, once provided the Fresnel coefficients, and is valid for \textit{any} material's frequency dispersion. In the low-loss limit, the emission spectrum is maximum at $\omega=\Omega$, where Eq.~\eqref{eq:Phi_analytical} reduces to the result by Miller \textit{et al}. in \cite{miller_shape-independent_2015}, as shown in Appendix \ref{appendixA}, where we discuss an approach to quantitatively distinguish the low-loss from the high-loss regime in NFRHT (Eq. \eqref{Seq:Qth}).
\begin{figure}[t]
\centering
\includegraphics[width=\columnwidth]{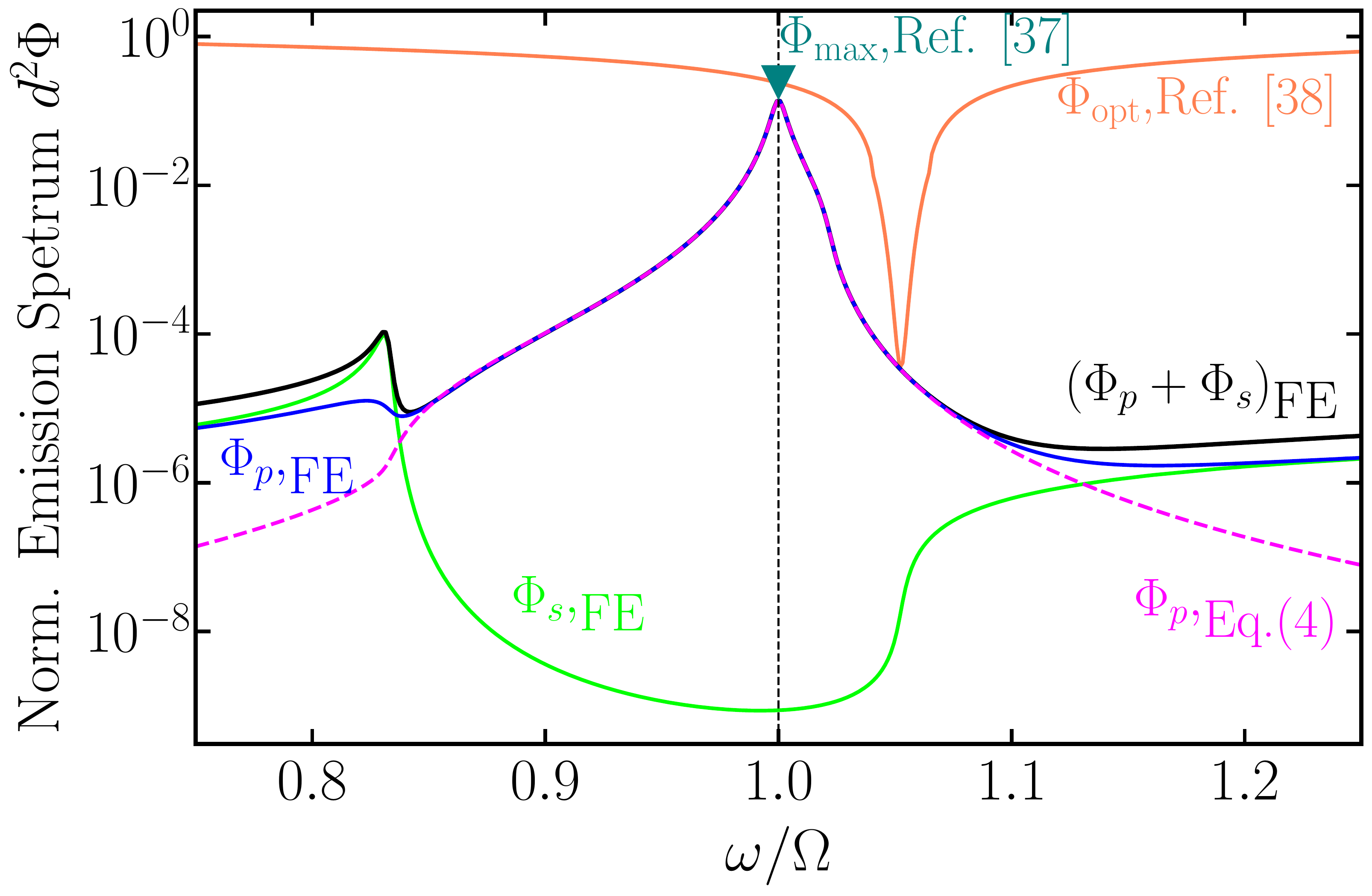}
\caption{Thermal emission spectrum, $\Phi$, normalized by $d^{-2}$ as a function of frequency, in the neighborhood of the SPhP resonance $\Omega$ for two bulk planar layers made of SiC exchanging thermal radiation in the near-field. $ \Phi_{p,\text{FE}}$, $ \Phi_{s,\text{FE}}$, and $ \Phi=(\Phi_{p}+\Phi_{p})_{\text{FE}}$, shown with the blue, green, and black curve, respectively, correspond to the p-polarization, s-polarization, and total spectrum as computed via fluctuational electrodynamics, whereas the analytical prediction in Eq. \eqref{eq:Phi_analytical} as well as in \cite{rousseau_asymptotic_2012} is shown with the magenta dashed line. We also show the spectral upper bound $\Phi_\text{opt}$ from \cite{venkataram_fundamental_2020} (orange curve) and the emission spectrum at the resonance frequency from \cite{miller_shape-independent_2015} (teal triangle), given in Eq. \eqref{Seq:MillerBound} of Appendix \ref{appendixA}.}
\label{fig:fig1_SiC}
\end{figure}

The dielectric function of {most} plasmonic and polar materials can be described by single Drude and Lorentz oscillators, respectively:
\begin{subequations}
  \label{eq:plasm_pol_dispers}
    \begin{empheq}[left={\varepsilon\bomega=\empheqlbrace\,}]{align}
        & \displaystyle\epsdrude  =\varepsilon_\infty \left[1-\frac{\omega_p^2}{\omega(\omega+i\,\gamma)}\right]\label{eq:eps_plasm}\\
         &\displaystyle\epsdl  = \varepsilon_\infty\left[1+\frac{\omega_\text{LO}^2-\omega_\text{TO}^2}{\omega_\text{TO}^2-\omega^2 -i\,\omega \gamma}\right],\label{eq:eps_pol}
    \end{empheq}
\end{subequations}
where $\varepsilon_\infty$ is the high-frequency relative permittivity, and $\gamma$ is the optical loss \cite{drude_zur_1900,kheirandish_modified_2020}. For plasmonic metals, $\omega_p$ is the plasma frequency, near which the SPP mode occurs. For polar materials, $\omega_\text{TO}$ and $\omega_\text{LO}$ are the transverse and longitudinal optical phonon frequencies, respectively \cite{caldwell_low-loss_2015}. The spectral range $[\omega_\text{TO},\omega_\text{LO}]$ defines the \emph{Reststrahlen} band \cite{kortum_phenomenological_1969}, within which the SPhP mode occurs.

The resonance frequency, $\Omega$, is found by solving Eq.~\eqref{eq:resonance_freq}. Since $\Omega$ does not vary significantly as $\gamma$ increases (see Appendix \ref{appendixC}), we evaluate it for $\gamma\to 0$ as: 

\begin{subequations}
  \label{eq:omega_res_gen}
    \begin{empheq}[left={\Omega=\empheqlbrace\,}]{align}
         &\displaystyle\Omegaplasm = \sqrt{\frac{\varepsilon_\infty}{\varepsilon_\infty+1}}\omega_p \label{eq:Omega_plasm}\\
         &\displaystyle  \Omegapolar = \sqrt{\frac{\varepsilon_\infty \omega_\text{LO}^2+\omega_\text{TO}^2}{1+\varepsilon_{\infty}}}\label{eq:Omega_pol}. 
    \end{empheq}
\end{subequations}
Henceforth, we assume that $\Omegaplasm$ and $\Omegapolar$ are $\gamma$-independent.

As an example, in Fig. \ref{fig:fig1_SiC}, we plot $\Phi(\omega)$ normalized with $d^{2}$ for silicon carbide (SiC), a widely used polar dielectric in the NFRHT literature \cite{song_near-field_2015,francoeur_electric_2011,mulet_enhanced_2002}. We consider a representative Lorentz model for its permittivity, with $\varepsilon_\infty=6.7$, $\omega_\text{TO}=1.49\times 10^{14}\rads$, $\omega_\text{LO}=1.83\times 10^{14}\rads$, $\gamma=8.97\times 10^{11}\rads$ \cite{hong_near-field_2018}. The exact emission spectrum for $p$-polarization ($\Phi_{p,\text{FE}}$), obtained via numerical integration with no approximations, is shown with the blue curve. Its $s$-polarization counterpart ($\Phi_{s,\text{FE}}$) as well as their sum, $\Phi = (\Phi_{p}+\Phi_{p})_{\text{FE}}$, are shown with the green and black curve, respectively. As anticipated, the $p$-polarization component dominates the emission spectrum in almost the entire frequency range near $\Omega$, and thus coincides with the total $\Phi$. Importantly, the magenta dashed curve shows $\Phi_p$ obtained with the analytical solution in Eq. \eqref{eq:Phi_analytical} and in \cite{rousseau_asymptotic_2012}. This curve overlaps nearly perfectly with FE for frequencies near-resonance. We also display an upper bound to the spectrum of NFRHT, $\Phi_\text{opt}$, as derived by Venkataram \textit{et al.} \cite{venkataram_fundamental_2020}, obtained through singular value decomposition of the Maxwell Green's tensor. This result accurately estimates the response of SiC only on resonance. Similarly, we show with a triangle-shaped marker the upper bound to NFRHT on resonance, derived by Miller \textit{et al.} \cite{miller_shape-independent_2015}, and given in Eq. \eqref{Seq:MillerBound}. 
\subsection{Optimal loss and upper bounds}
In evaluating NFRHT performance, it is useful to introduce a material quality factor for the polaritonic resonance in plasmonic and polar media \cite{wang_general_2006,pascale_bandwidth_2021,caldwell_low-loss_2015}: 
\begin{equation}\label{eq:Q}
\displaystyle    Q=\left.\frac{\omega \frac{d\real{\varepsilon}}{d\omega}}{2\imag{\varepsilon}}\right|_\Omega\approx \frac{\Omega}{\gamma}.
\end{equation}

To obtain the radiative thermal conductance (Eq.~\eqref{eq:HTC_def}), we use contour integration in the complex frequency plane of $\Phi$ (see Appendices \ref{appendixB} and \ref{appendixC}), similar to \cite{rousseau_asymptotic_2012}. Considering that the Planck distribution varies slowly with respect to $\Phi\bomega$, we obtain:
\begin{equation}\label{eq:HTC_analytical}
    h = \hmax \, \Psi\left(\frac{Q}{B}\right)\PiT .
\end{equation}
The functions $\Pi$ and $\Psi$ are given by:
\begin{equation}\label{eq:PiT}
    \quad\displaystyle \PiT=\frac{1}{k_B}\frac{\partial}{\partial T}\theta(\Omega,T) =\left[\frac{\frac{\hbar }{2k_B} \frac{\Omega}{T}}{\sinh{\left(\frac{\hbar }{2k_B} \frac{\Omega}{T}\right)}}\right]^2 ,
\end{equation}
\begin{equation}\label{eq:Psi}
\displaystyle
        \Psi\left(\frac{Q}{B}\right)=\left.\frac{h}{\hmax}\right|_{T\gg \frac{\hbar \Omega}{2k_B}} = -\frac{\Li{-(Q/B)^2}}{1.36 (Q/B)}.
\end{equation}
These functions are both bounded above by unity, hence $\hmax$ in Eq.~\eqref{eq:HTC_analytical} defines the maximum thermal conductance that a polaritonic material can reach in a planar configuration, and is given by:
\begin{equation}\label{eq:hmax_z}
    \hmax = \frac{1.36 \, k_B}{16\pi d^2} \frac{\Omega}{B}.
\end{equation}
As can be seen, $\hmax$ is temperature- and loss-independent. Furthermore, it expresses the well-known $\propto d^{-2}$ dependence of NFRHT from the vacuum gap size \cite{wang_parametric_2009,ben-abdallah_fundamental_2010,rousseau_asymptotic_2012,iizuka_analytical_2015}. 
It must be noted that this result is valid within a macroscopic description of thermal fluctuations. To properly estimate how NFRHT scales in the limit $d\to 0$, often referred to as the "extreme near-field", one should consider effects relevant at microscopic scales, e.g., transport properties of the materials a well as non-local electromagnetic response \cite{chapuis_effects_2008,esfarjani_heat_2011,chiloyan_transition_2015,venkataram_phonon-polariton_2018}. 

The parameter $B$ is termed \emph{material residue} henceforth, and it is defined as $\displaystyle B =  Q/\imag{r_p(\Omega)}$, by setting $Q\to \infty$, which reduces to:
\begin{subequations}
\label{eq:Bplasm_pol}
  \begin{align}[left={B = \empheqlbrace}]
    & \displaystyle\Bplasm = \frac{1+\varepsilon_\infty}{2}\label{eq:Bplasm}\\
    &  \displaystyle  \Bpolar=\frac{(1+\varepsilon_\infty)^2}{2\varepsilon_\infty}\frac{\Omegapolar^2}{\omega_\text{LO}^2-\omega_\text{TO}^2}\label{eq:Bpol},           
   \end{align}
  \end{subequations}
for Drude and Lorentz materials, respectively.

Eq.~\eqref{eq:HTC_analytical} is the key contribution of this paper. Unlike expressions presented in previous works \cite{rousseau_asymptotic_2012,iizuka_analytical_2015}, Eq.~\eqref{eq:HTC_analytical} distinctly separates the role of the optical loss, described by the quality factor $Q$, in NFRHT, from other dispersion parameters that are captured by $B$, and temperature. 
The decoupling of temperature, material quality factor, and material residue, in Eq.~\eqref{eq:HTC_analytical}, via the functions $\displaystyle \Pi\left(\frac{\Omega}{T}\right)$ and $\displaystyle\Psi\left(\frac{Q}{B}\right)$, allows a quantitative classification of different materials as candidates for tailoring NFRHT. 

{Eq.~\eqref{eq:HTC_analytical} is a very good approximation of the exact result obtained via fluctuational electrodynamics, for $d\ll \lambda_\mathrm{T}$ and $Q$ considerably larger than unity, which is satisfied by all relevant materials for NFRHT} \footnote{As a benchmark, for $Q\gg 1$, considered in \cite{miller_shape-independent_2015,venkataram_fundamental_2020}, the ratio $\displaystyle \frac{Q}{B}$ can be written as $\displaystyle \frac{Q}{B}=\frac{|\chi(\Omega)|^2}{\imag{\chi (\Omega)}}=\zeta$, where $\zeta$ is a material response factor}. Further, Eq.~\eqref{eq:HTC_analytical} is \textit{exact} for $\varepsilon_\mathrm{\infty}=1$ in either polar or plasmonic media. We stress that Eq.~\eqref{eq:hmax_z} represents a tight bound to NFRHT that accounts for material dispersion. This is to be contrasted to previous results that derived upper bounds to $h$ with the idealized assumption of a dispersion-less perfect blackbody in the near-field (i.e., $\xi=1$) \cite{ben-abdallah_fundamental_2010}, thus yielding a thermal conductance that is orders of magnitude larger than our result in Eq.~\eqref{eq:HTC_analytical} (Fig.~\ref{fig:fig3} (b)).

\begin{figure}[t]
\centering
\includegraphics[width=\columnwidth]{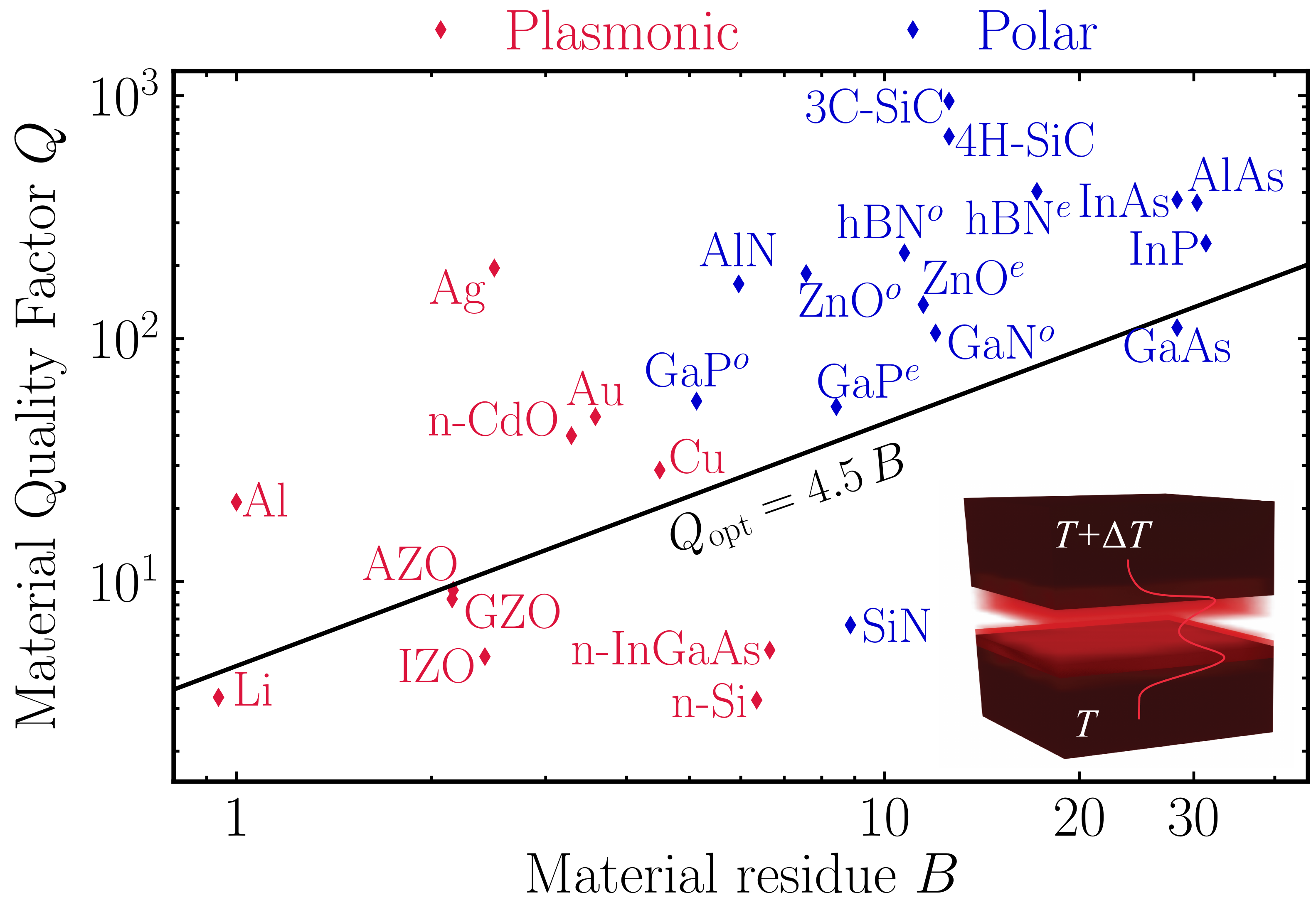}
\caption{{Materials' quality factor $Q$ and residue parameter $B$ for plasmonic and polar materials. Drude parameters are taken from \cite{ashcroft_solid_1976} for Au, Ag, Cu, Al. Lorentz parametetrs for SiN are taken from \cite{cataldo_infrared_2012} (upon fitting),  for n-doped Si from \cite{basu_infrared_2009}, whereas  for the rest of the considered materials from \cite{caldwell_low-loss_2015}. All doped semiconductors' dispersions are assumed independent from temperature. Superscripts ${}^o$ and ${}^e$ stand for the ordinary and extraordinary principal axes, respectively. The solid line shows Eq. \ref{eq:Qopt}, which maximizes NFRHT.}}
\label{fig:fig1}
\end{figure}

With Eqs. \eqref{eq:HTC_analytical}, \eqref{eq:Psi}, one can identify the optimal material characteristics, independent of temperature, that maximize NFRHT. In particular, $\Psi$ describes how NFRHT changes with optical loss. Seeking for the maximum of $\Psi$ (Eq.~\eqref{eq:Psi}), we obtain:
\begin{equation}\label{eq:Qopt}
    \Qopt=4.5 \,B.
\end{equation}
Hence, NFRHT is maximized when the material quality factor is $4.5$ times the material residue function, given in Eq.~\eqref{eq:Bplasm_pol}. The work in \cite{ben-abdallah_fundamental_2010} yielded a universal optimal quality factor, namely $Q_\mathrm{opt}^{*}=2.72$, which, however, is independent of $B$, thus suggesting that \textit{all} materials that have the same $Q$ should perform identically in terms of NFRHT. %This result is shown with the dashed line in Fig.~\ref{fig:fig1}. 
In contrast, Eq.~\eqref{eq:Qopt} demonstrates that other dispersion characteristics, beyond the quality factor, are critical in evaluating NFRHT response (see Appendix \ref{appendixC}).

From Eq.~\eqref{eq:Bplasm}, the material residue for plasmonic materials depends only on $\varepsilon_\infty$. Typically, $\varepsilon_\infty\lesssim 10$, hence $\Bplasm$ remains well below $10$. Thus, from Eq.~\eqref{eq:Qopt}, $\Qopt$ for plasmonic materials is relatively low, namely $\Qopt\lesssim 50$. Hence, plasmonic materials with good NFRHT performance have high-loss ($\gamma$) and modest $Q$, and NFRHT is enhanced due to the broadband nature of the plasmonic resonance. In contrast, the material residue for polar materials, $\Bpolar$ (Eq.~\eqref{eq:Bpol}) is inversely proportional to the spectral width of the Reststrahlen band, $(\omega_\mathrm{LO}-\omega_\mathrm{TO})$. The Reststrahlen band of most polar materials is narrow, hence $\Bpolar\gtrsim\Bplasm$, therefore $\Qopt$ for polar media is higher than for plasmonic ones. In contrast to plasmonic media, in polar ones, it is the narrowband nature of SPhPs that enhances NFRHT.

In Fig.~\ref{fig:fig1}, Eq.~\eqref{eq:Qopt} is shown with the solid line. We also evaluate the NFRHT performance of several relevant polaritonic emitters considered in literature. These include polar materials such as Silicon Carbide (SiC), hexagonal Boron Nitride (hBN), and doped semiconductors, e.g., Gallium Arsenide (GaAs), Indium Arsenide (InAs) \cite{cardona_fundamentals_2005,schubert_infrared_2000,caldwell_low-loss_2015}, as well as plasmonic materials such as standard noble metals, e.g., Gold (Au), Silver (Ag), and heavily doped oxides, e.g., IZO and GZO \cite{kim_optimization_2013,kim_plasmonic_2013,caldwell_low-loss_2015}. 
{ Although the presented framework models NFRHT for materials with only a single polaritonic resonance, i.e., one oscillator in the dielectric permittivity function, in section \ref{sec:multipleRes} we provide a semi-analytical extension able to describe also materials with multiple polaritonic resonances, e.g., SiO$_2$ \cite{chen_extraordinary_2007} or Al$_2$O$_3$ \cite{rajab_broadband_2008}. This model is in very good agreement with FE calculations, as long as the polaritonic resonances are spectrally sufficiently distant, as we show for the case of SiO$_2$.
} 

The distance between each point in Fig.~\ref{fig:fig1} and the solid curve representing Eq.~\eqref{eq:Qopt} expresses how far from the ideal material performance each material falls. Interestingly, an ultra-high $Q$ does \textit{not} necessarily yield optimal NFRHT. By contrast, it is the interplay between $Q$ and $B$ that is critical, making, for instance, GaAs, AZO and GaN near-optimal materials for NFRHT as compared to Ag or 3C-SiC, even though the latter exhibit ultra-high quality factors. This demonstrates the importance of the material residue, $B$, in evaluating NFRHT performance. 

\begin{figure}[t]
\centering
\includegraphics[width=\columnwidth]{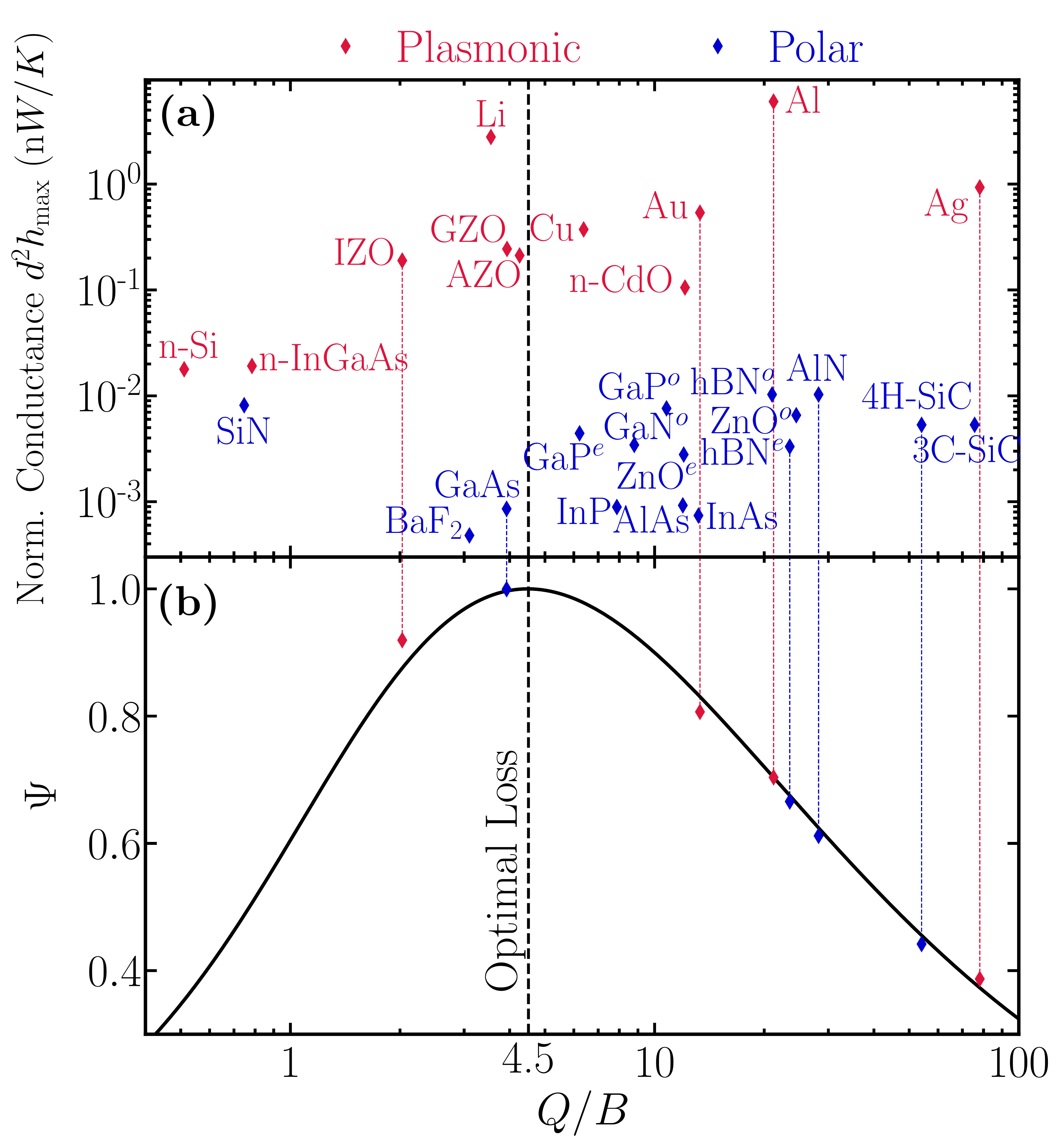}
\caption{{\bf(a)} Maximum heat transfer coefficient, $\hmax$ (Eq.~\eqref{eq:hmax_z}), for pairs of plasmonic and polar materials as considered in Fig.~\ref{fig:fig1}, for optimal loss and infinite temperature. {\bf(b)} Function $\Psi$ (Eq.~\eqref{eq:Psi}), as $Q/B$ varies. The optimal $\Qopt/B$ is shown with the dashed vertical line. Points in panel (b) represent fluctuational electrodynamics calculations for the considered materials, and are in very good agreement with our analytical result (Eq.~\eqref{eq:HTC_analytical}).}
\label{fig:fig2}
\end{figure}

The parameter $\hmax$ in Eq.~\eqref{eq:hmax_z} is the maximum radiative thermal conductance achievable for each material, if one adjusted their quality factor such that $\displaystyle\Psi\left(\frac{Q}{B}\right)\to 1$, and in the limit of infinite temperature, for which $\displaystyle\PiT\to 1$. In Fig.~\ref{fig:fig2} (a), we calculate $\hmax$ for the materials considered in Fig.~\ref{fig:fig1}. To quantify the degree to which the loss of each material deviates from the optimal value as defined in Eq.~\eqref{eq:Qopt}, we plot these points against the ratio $Q/B$, which is inversely proportional to $\gamma$, the optical loss. Fig.~\ref{fig:fig2} (a) demonstrates that materials with significantly different quality factors, e.g., Au and Ag, can have similar maximal thermal conductance, if one adjusted their loss. This occurs because the material residue, $B$, compensates for the lower $Q$ of Au as compared to that of Ag (see Fig.~\ref{fig:fig1}). In other words, small deviations of $\varepsilon_\mathrm{\infty}$ from unity in plasmonic metals (Eq.~\eqref{eq:Bplasm}), and, similarly, sub-optimal Reststrahlen band spectral widths with respect to $\Omega$ in polar materials (Eq.~\eqref{eq:Bpol}), can considerably affect the optimal point of NFRHT. 

The dependence of NFRHT from the optical loss is captured explicitly in $\Psi$ (Eq.~\eqref{eq:Psi}), and is shown graphically in Fig.~\ref{fig:fig2} (b). As described in Eq.~\eqref{eq:Qopt}, $\Psi$ is maximum at $\Qopt=4.5\,B$, depicted with the vertical dashed line. The horizontal distance between this line and each point in Fig.~\ref{fig:fig2} (a) indicates how close each material is to the ideal optical loss, for its particular resonance frequency, $\Omega$. For example, despite the comparable material residue, $B$, of Ag and Au, the loss ($\gamma$) of Au yields a value of $\Psi$ that is much closer to unity as compared to Ag, hence Au presents overall better NFRHT performance, which is consistent with Fig.~\ref{fig:fig1}.

In Fig.~\ref{fig:fig2} (b), we also append points that correspond to exact calculations with fluctuational electrodynamics, for few commonly used materials in NFRHT. These calculations are performed in the limit of infinite temperature, for the sake of a meaningful comparison with our formalism in Eq.~\eqref{eq:HTC_analytical}. It is important to stress that the limit of infinite temperature here has only the mathematical purpose of saturating the function $\Pi$, and therefore remove the temperature dependence in $h$.
The exact results, represented as markers, are in very good agreement with our theory (solid line) for all considered materials. Small discrepancies occur in the range of relatively low-$Q$, for example in the case of IZO \cite{kim_plasmonic_2013}, since our formalism assumes resonant material response, hence its accuracy improves as the material quality factor increases (see Appendix \ref{appendixB} for details).
\subsection{Temperature dependence}
\begin{figure}[htbp]
    \centering
    \includegraphics[width=\columnwidth]{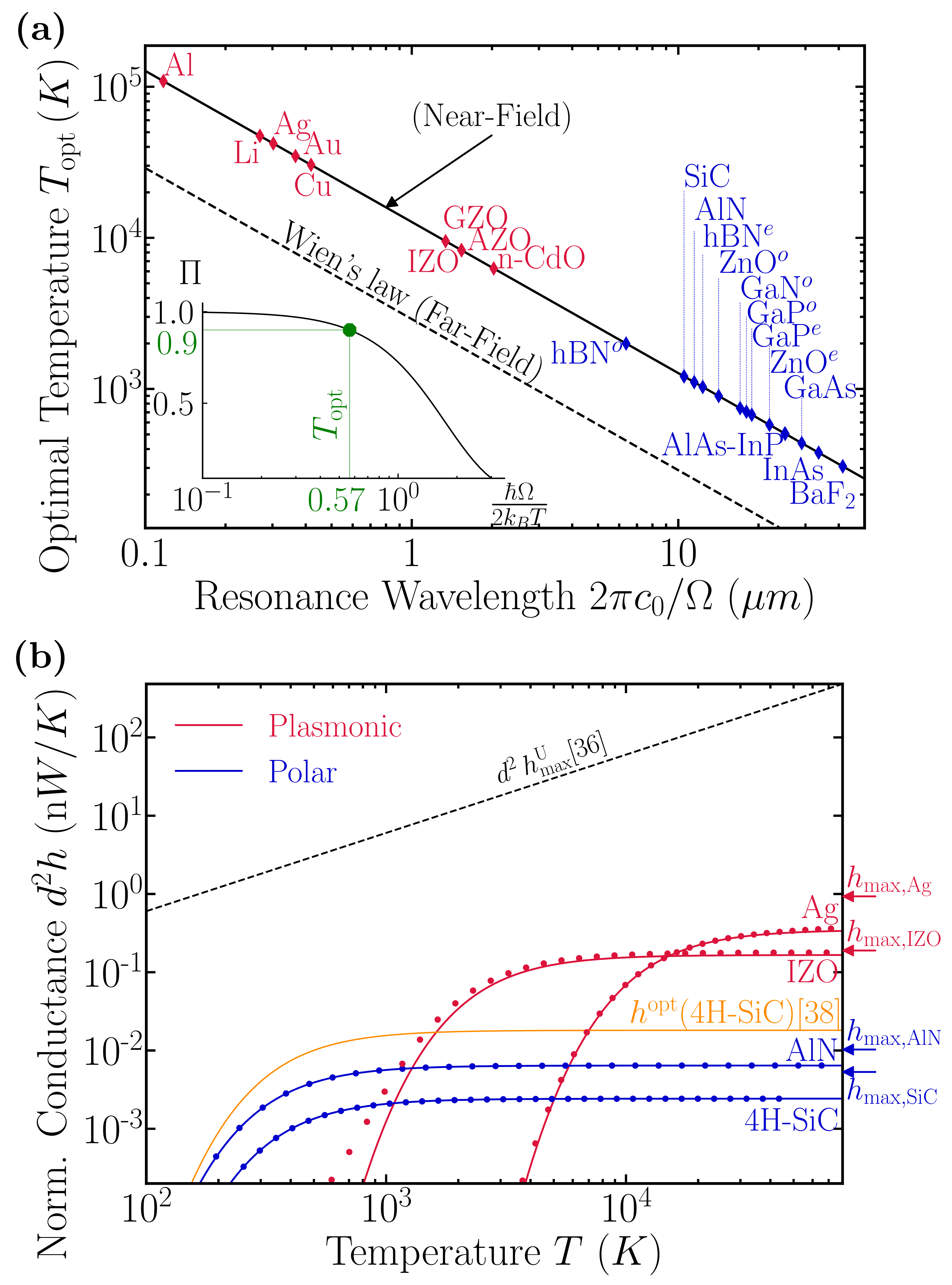}
    \caption{{\bf (a)} Optimal temperature $\Topt$ as a function of the resonance wavelength, calculated using Eq. \eqref{Seq:Topt}. 
    The Drude parameters for the Au, Ag, Cu, Al are taken from \cite{ashcroft_solid_1976}, while the Drude parameter for the other plasmonic materials and the Lorentz parameters for polar materials are taken from \cite{caldwell_low-loss_2015} (tables 1-2). The superscript ${}^o$ and ${}^e$ stand for the ordinary and extraordinary principal axes of the corresponding material, respectively.
    In the inset, we plot $\Pi$ (in Eq. \eqref{eq:PiT}), which expresses the normalized  thermal conductance in the optimal loss condition $Q=\Qopt=4.5\,B$ (see Eq. \eqref{eq:Qopt}). We identify the optimal operating temperature for a material by setting $\Pi=0.9$ (green marker). {\bf (b)}Temperature dependence of the total radiative thermal conductance, $h$ (Eq.~\eqref{eq:HTC_analytical}), for pairs of plasmonic materials (Ag, IZO) and polar ones (AlN, 4H-SiC). Dotted lines show results with fluctuational electrodynamics. {the fundamental bound $ \hmax^\mathrm{U} = \frac{k_B^2 T}{3\hbar d^2}$ \cite{ben-abdallah_fundamental_2010} is shown with the black dashed line and the upper bound $h^\text{opt}$ by \cite{venkataram_fundamental_2020} is shown with the orange curve for 4H-SiC.}}
    \label{fig:fig3}
\end{figure}

{%
The temperature dependence of NFRHT is described via $\displaystyle\PiT$ in Eq.~\eqref{eq:PiT}, displayed in the inset of Fig. \ref{fig:fig3} (a),  which is the only temperature-dependent term in Eq.~\eqref{eq:HTC_analytical}, and agrees with previous analytical results \cite{ben-abdallah_fundamental_2010,rousseau_asymptotic_2012}. 
In contrast to Wein's displacement law in the far-field, where $h$ scales as $T^{3}$, in the near-field, $\Pi$ scales as $T^{-2}$ (for increasing $T$).
On the other hand, from Eq.\eqref{eq:hmax_z}, $\hmax$ scales with the resonance frequency, therefore materials supporting polaritons at high frequencies (high-$\Omega$) will, in principle, reach higher NFRHT rates. However, for this to occur in practice, they ought to operate at dramatically higher temperatures. Specifically, since $\Pi$ decreases exponentially with the ratio $\Omega/T$, to avoid a dramatic damping in $h$, a higher resonance frequency should be compensated by a higher operating temperature, as expected. 

This is well-understood in the far-field regime with Wien's displacement law that estimates the optimal resonance frequency of a thermal emitter at a given temperature for maximizing the power emitted in the far-field. One can similarly estimate the optimal temperature, $T_\mathrm{opt}$, of a polaritonic thermal emitter in a planar near-field configuration, by maximizing $\displaystyle\PiT$. Since $\Pi$ reaches its maximum $\Pi=1$ in the limit of infinite temperature, we compute the optimal temperature of operation as a function of resonance frequency, $\Omega$, by setting the term $\displaystyle\PiT$ to $0.9$, as shown with the green lines in the inset of Fig. \ref{fig:fig3} (a). The resulting optimal temperature is expressed as:
\begin{equation}\label{Seq:Topt}
        \Topt = \frac{\hbar \Omega}{2k_B 0.57} = \frac{\bNF}{\lambda},
\end{equation}
where $\bNF= 12729\mu m \, K\approx 4.4\times  \bWien$ (in which the subscript stands for Near Field) and $\lambda =2\pi c_0/\Omega $ is the resonance wavelength. This dependence of $T_\mathrm{opt}$ from $\Omega$ is shown with the solid line in Fig. \ref{fig:fig3} (a). As a reference, we also display with the dashed line Wein's displacement law, relevant in the far-field. One can therefore see that in the near-field, considerably higher temperatures are required to reach optimal material performance, as compared to the far-field. This is expected, since far-field thermal emission is generally more broadband in comparison to near-field thermal emission, especially for polaritonic materials \cite{joulain_surface_2005,song_near-field_2015}. Hence, the maximal spectral overlap between the mean energy per photon, $\theta(\omega,T)$, and a blackbody spectrum is achieved at much lower temperatures as compared to the overlap between a narrowband near-field resonance and $\theta(\omega,T)$, since $\theta$ broadens as $T$ increases.

In Fig. \ref{fig:fig3} (a) we also display the optimal temperature, calculated using Eq. \eqref{Seq:Topt}, as a function of the resonance wavelength $2\pi c_\mathrm{o}/\Omega$, for the polar and plasmonic materials considered in Figs. \ref{fig:fig1}-\ref{fig:fig2}. It can be seen that most polar media, with resonance frequencies mainly in the mid-IR, will achieve optimal performance at temperatures that are up to two orders of magnitude lower than their plasmonic counterparts, since plasmonic resonances occur mainly in the near-IR, visible and UV regimes. }

This can also be seen in Fig.~\ref{fig:fig3} (b), showing the total radiative thermal conductance, $h$, computed via our analytical result (Eq.~\eqref{eq:HTC_analytical}). We consider a set of plasmonic materials, i.e., IZO and  Ag, and a set of polar ones, i.e., 4H-SiC and AlN. It is clear that plasmonic materials reach higher NFRHT than their polar counterparts, however this occurs at very high temperatures. This is expected since the resonance frequency, $\Omega$, of plasmonic media is significantly higher than that of polar ones. 

In Fig.~\ref{fig:fig3} (b), we also append the exact results with fluctuational electrodynamics (dotted), where the wavenumber and frequency integrations are carried out numerically. These are in excellent agreement with our analytical formalism, except for small deviations that occur only for materials with relatively low $Q$. This is expected, since a low-$Q$ suggests a spectrally broadband response, whereas our formalism applies to polaritonic resonances (Fig. \ref{fig:fig3S}(a) of Appendix \ref{appendixC}). To conclude, the vast majority of polar and plasmonic materials, one can compute \textit{exactly} their NFRHT properties with Eqs. (\ref{eq:HTC_analytical}-\ref{eq:hmax_z}). 

{ As a reference, in Fig.~\ref{fig:fig3} (b), we also show the fundamental limit to the radiative thermal conductance $h$, i.e., $\displaystyle \hmax^\mathrm{U} = \frac{k_B^2 T}{3\hbar d^2}$, as derived by Ben-Abdallah \etal \cite{ben-abdallah_fundamental_2010} (dashed line), and the upper bound derived by Venkataram \etal \cite{venkataram_fundamental_2020} for one of the considered materials, viz. 4H-SiC, denoted with $h^\text{opt}$(4H-SiC) (orange curve). By comparing with our exact results, both $\hmax^\mathrm{U}$ and $h^\text{opt}$(4H-SiC) represent loose bounds to NFRHT. This is to be contrasted with the expression in Eq.~\eqref{eq:hmax_z}, which is the limit to which $h$ actually saturates at high temperatures for optimal loss, i.e., $Q=\Qopt$, for every material (see right $y$-axis in Fig.~\ref{fig:fig3} (b)).}

\subsection{Polar dielectrics with multiple polaritonic resonances}\label{sec:multipleRes}
Beyond the theory presented in the previous sections, there exist several polaritonic materials, for example SiO$_\mathrm{2}$, Al$_2$O$_3$ and MoO$_\mathrm{3}$  that exhibit multiple polaritonic resonances within the IR range \cite{chen_extraordinary_2007,rajab_broadband_2008,ma_-plane_2018}. In this section, we expand upon the results of the previous sections and present a semi-analytical method to include multiple polaritonic resonances. 

Let us consider the dielectric permittivity $\varepsilon\bomega$ of the bulk layers as sum of $N$ Drude or Lorentz oscillators, i.e.,
$
    \varepsilon\bomega = \sum_{n=1}^N \varepsilon_n\bomega,
$
where $\varepsilon_n\bomega$ is given in Eq. \eqref{eq:plasm_pol_dispers}. The polaritonic resonance of each oscillator is assumed spectrally far from each other.
In this case, a good approximation of the total emission spectrum $\Phi\bomega$ is given by a superposition of the emission spectra of the single oscillators: 
\begin{equation}\label{Seq:Phi_sum}
    \Phi\bomega = \sum_{n=1}^N \alpha_n \Phi_n\bomega,
\end{equation}
where $\Phi_n\bomega$ is calculated by inserting the single oscillator permittivity $\varepsilon_n\bomega$ in Eq. \eqref{eq:Phi_analytical}, and $\alpha_n$ is calculated numerically by fitting $\Phi\bomega$ with the exact numerical FE result. 
Assuming the function $\PiT$ slowly varying with respect to each emission spectrum $\Phi_n$, which peaks at the polariton resonance frequency $\Omega_n$, we can calculate the radiative thermal conductance as superposition of the single oscillator conductances, i.e., 
\begin{equation}\label{Seq:h_sum}
    h=\sum_{n=1}^N \alpha_n h_n,
\end{equation}
where $h_n$ is obtained by plugging each oscillator's parameters, i.e., $\{\Omega_n,\, Q_n,\,B_n\}$, into Eq. \eqref{eq:HTC_analytical}. Therefore, each oscillator has an optimal quality factor $Q_{\text{opt},n}=4.5\,B_n$, with the material residue $B_n$ given in Eq. \eqref{eq:Bplasm_pol}, and an upper bound $h_{\text{max},n}$, given in Eq. \eqref{eq:hmax_z}. The total conductance upper bound $\hmax$ will be given by the linear combination of each oscillator upper bound, i.e. $    \hmax=\sum_{n=1}^N \alpha_n h_{\text{max},n}$.

\begin{figure}[h!]
    \centering
    \includegraphics[width=\columnwidth]{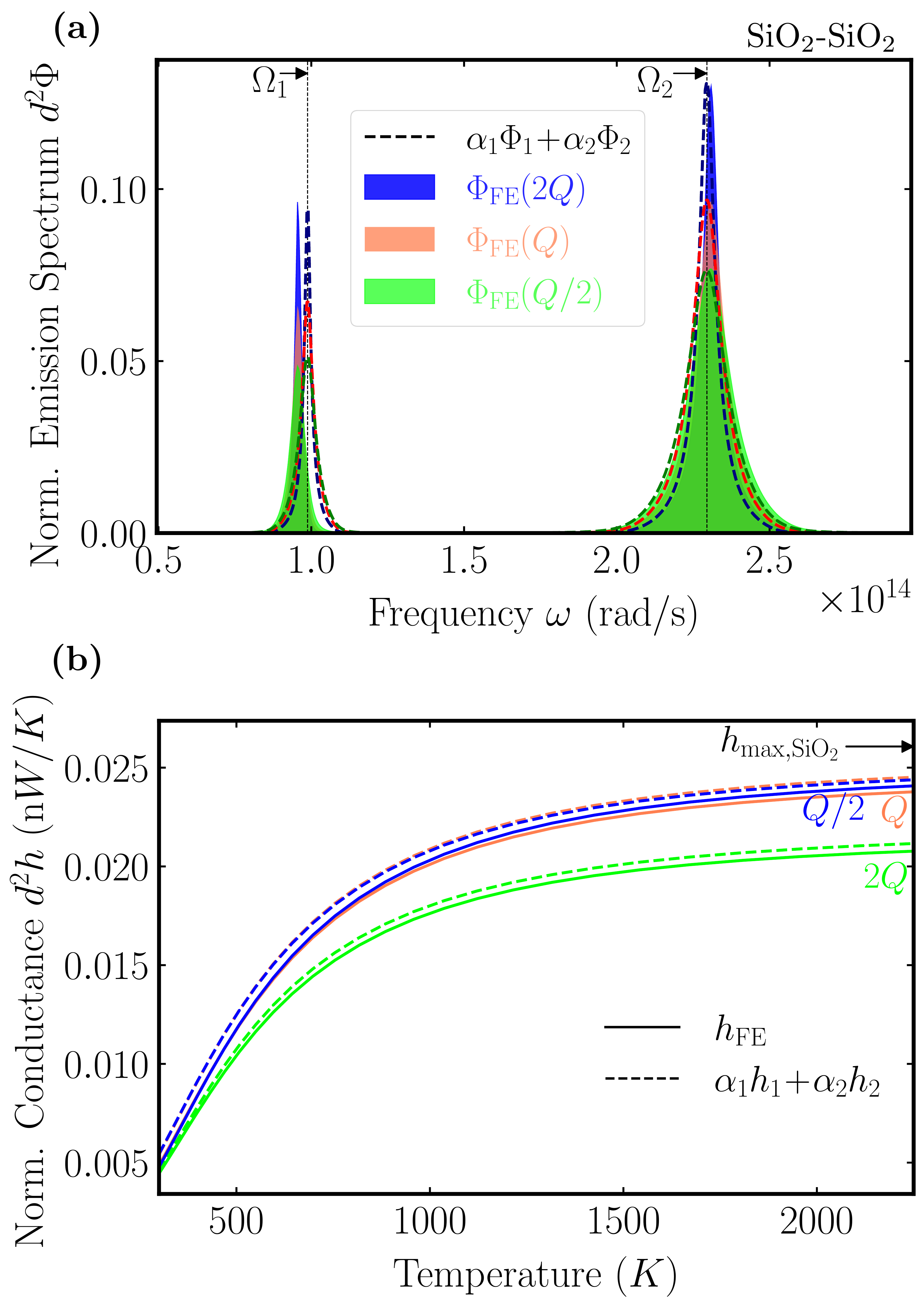}
    \caption{{{\bf (a)} Thermal emission spectrum, $\Phi$, normalized with $d^{2}$ as a function of frequency, for two bulk planar layers of SiO$_2$ exchanging thermal radiation in the near-field. The dispersion relation is the superposition of two Lorentz oscillators, with dispersion parameters $\varepsilon_\infty=1.007$, $\omega_{TO,1}=8.6734\times10^{13} \rads$, $\omega_{TO,2}=2.0219\times10^{14} \rads$,  $\omega_{LO,1}=1.0953\times10^{14} \rads$,  $\omega_{LO,2}=2.5387\times10^{14} \rads$,  $\gamma_{1}=3.3026\times10^{12} \rads$, $\gamma_{2}=8.3983\times10^{12}, \rads$ \cite{chen_extraordinary_2007}, and resonance frequencies $\Omega_1 = 9.8793 \times 10^{13} \rads$ and $\Omega_2 = 2.2950 \times 10^{14} \rads$ . The shaded areas correspond to fluctuational-electrodynamics (FE) calculations, whereas the analytical prediction in Eq. \eqref{Seq:Phi_sum}, with $\alpha_1=0.7,\alpha_2=1$ is shown with dashed lines. We plot $\Phi$ for three values of optical loss, viz., of $\gamma_i,$ or equivalently $Q_i=\Omega_i/\gamma_i$ ($i=1,2$): $\gamma_i$ ($Q_i$, in orange), $\gamma_i\to2\gamma_i$ ($Q_i\to Q_i/2$, in green), $\gamma_i\to \gamma_i/2$ ($Q_i\to 2Q_i$, in blue). In {\bf (b)}, we plot the normalized radiative heat conductance $h$ for the cases considered in panel (a). Solid lines correspond to FE calculations, whereas dashed lines to the analytical prediction in Eq. \eqref{Seq:h_sum}. The total upper bound calculated as $h_{\text{max,SiO}_2}=\alpha_1h_{\text{max},1}+\alpha_2h_{\text{max},2}$ is also shown.}}
    \label{fig:SiO2}
\end{figure}  

We apply this approach to describe the radiative heat conductance of two bulk SiO$_2$ layers, with permittivity given by the sum of two Lorentz oscillators whose parameters, taken from \cite{chen_extraordinary_2007}, are given in the figure's caption. By plugging these parameters in Eq. \eqref{eq:Omega_pol}, we calculate the SPhP resonance frequencies, i.e., $\Omega_1 = 9.8793 \times 10^{13} \rads$ and $\Omega_2 = 2.2950 \times 10^{14} \rads$. In Fig. \ref{fig:SiO2} (a), we plot the total spectrum $\Phi_{FE}$ computed via fluctuational electrodynamics, including s- and p-polarization contributions for $d=10\,$nm, and the spectrum $\Phi$ calculated as sum of the single oscillator contributions. We calculate also the emission spectrum in two other loss scenarios, i.e., $\gamma_i\to2\gamma_i$ and $\gamma_i\to\gamma_i/2$ ($i=1,2$), corresponding to half or twice the quality factor of the single polariton resonances, respectively, according to the definition in Eq. \eqref{eq:Q}. 
By setting $\alpha_1=0.7$ and $\alpha_2=1$, it is evident that there is good agreement between the exact emission spectrum and the one reconstructed using Eq. \eqref{Seq:Phi_sum} in all considered loss scenarios, with a slight mismatch in the low-frequency resonance contribution, due to the influence of the high-frequency resonance contribution, not taken into account in our model. This agreement is preserved in the corresponding radiative heat conductance $h$, plotted in Fig. \ref{fig:SiO2} (b) as a function of the temperature. The total upper bound $h_{\text{max,SiO}_2}$, calculated as $h_{\text{max,SiO}_2}=\alpha_1h_{\text{max},1}+\alpha_2h_{\text{max},2}$ is also shown. 
\section{Conclusions}
We present a simple analytical framework that describes NFRHT in bulk systems with single polaritonic resonance. We derive a universal closed form expression (Eq.~\eqref{eq:HTC_analytical}) for the thermal conductance that is valid for {most plasmonic or polar materials}. This expression clarifies what the role of optical loss ($\gamma$) and material quality factor ($Q\propto\gamma^{-1}$) are in NFRHT, as well as their interplay with other material dispersion characteristics. We show that the quality factor of a material's polariton resonance alone is not sufficient to accurately describe NFRHT. In contrast, we introduced the material residue parameter, $B$, that completes the analytical framework for the classification of \text{all} plasmonic and polar materials for NFRHT.

We derive a material-dependent optimal condition that maximizes NFRHT, namely $Q=4.5\,B$, where the quality factor of the polaritonic resonance is inversely proportional to the optical loss, and the material residue is loss-independent and encompasses critical properties in polaritonic materials, i.e., the resonance frequency and the spectral width of the Reststrahlen band.

{Although previous works have derived upper bounds to the spectral emissivity \cite{shim_fundamental_2019,molesky_fundamental_2020,venkataram_fundamental_2020} and loose upper bounds to the total near-field thermal conductance \cite{pendry_radiative_1999,ben-abdallah_fundamental_2010}, here, we provide a \textit{tight}  bound to the thermal conductance, $\hmax$.} Other than the well-known dependence from the gap-size $d^{-2}$, $\hmax$ also rigorously demonstrates the role of other material dispersion characteristics in NFRHT. Finally, we provided a semi-analytical extension to our approach, able to describe multiple polaritonic resonances.

\section{Acknowledgments}
This work is dedicated to the memory of John S. Papadakis.
The authors acknowledge Dr. Mitradeep Sarkar for creating the schematic drawing in Fig.~\ref{fig:fig1}.
The authors declare no competing financial interest. G. T. P. acknowledges funding from ”la Caixa” Foundation (ID 100010434), from the PID2021-125441OA-I00 project funded by MCIN /AEI /10.13039/501100011033 / FEDER, UE, and from the European Union’s Horizon 2020 research and innovation programme under the Marie Sklodowska-Curie grant agreement No 847648. The fellowship code is LCF/BQ/PI21/11830019. This work is part of the R\&D project CEX2019-000910-S, funded by MCIN/ AEI/10.13039/501100011033/ , from Fundació Cellex, Fundació Mir-Puig, and from Generalitat de Catalunya through the CERCA program.

\appendix

\section{Derivation of Eq. \eqref{eq:Phi_analytical}}\label{appendixA}
We derive the analytical expression of the emission spectrum $\Phi$ for two semi-infinite layers made of non-magnetic, isotropic, homogeneous polaritonic material with relative dielectric permittivity $\varepsilon(\omega)$ and susceptibility $\chi\bomega =\varepsilon\bomega -1$,  separated by a vacuum gap of size $d$.

For gap sizes smaller than the thermal wavelength, evanescent waves ($\beta>k_0$) dominate the heat flux. 
In this range, the corresponding transmission probability, for $p$-polarization, can be expressed as \cite{rytov_theory_1953,polder_theory_1971,loomis_theory_1994,shchegrov_near-field_2000}:
$
   \displaystyle \xi_{p}(\omega,\beta>k_0) = \frac{4\imag{r_{p}}^2 e^{-2\eta_0 d}}{|1-r_{p}^2 e^{-2\eta_0 d}|^2},
$
where $\eta_0=\sqrt{\beta^2-k_0^2}$ is the out-of-plane wavenumber in vacuum, and $r_{p}$ is the Fresnel coefficients at the vacuum-material interface for $p$ \cite{yeh_optical_1988}. At sufficiently small vacuum gaps, the dispersion of surface polaritons approaches the quasistatic limit, for which $\beta\gg k_0$ \cite{pendry_radiative_1999}. In this limit, $\eta_0\approx\beta$, and one can approximate the Fresnel coefficient with 
$    r_p\bomega=\frac{\varepsilon\bomega-1}{\varepsilon\bomega+1}$.

The transmission probability $\xi_p$ for $p$-polarization in the electrostatic limit $\beta\gg k_0$ can be written as:
\begin{equation}\label{Seq:xi_p}
    \xi_{p}(\omega,x) = \frac{4\imag{r_p\bomega}^2 e^{-2 x}}{|1-r_{p}\bomega^2 e^{-2x}|^2},
\end{equation}

Perfect photon tunneling occurs at $\xi_p(\omega,\beta)=1$. From Eq. \eqref{Seq:xi_p}, this occurs at an in-plane wavenumber of:
\begin{equation}\label{Seq:beta_xi1}
    \beta_{res}\bomega  = \frac{1}{2d}\ln \left|r_p \bomega \right|^2 = \frac{1}{2d}\ln \left|\frac{\chi\bomega }{\chi\bomega  +2}\right|^2.
\end{equation}
Eq. \eqref{Seq:beta_xi1} is valid for frequencies $\omega$ such that $\left|1+2/\chi\bomega \right|<1$, and defines a curve in the $(\omega,\beta)$ parameter space, near which the NFRHT is maximal \cite{ben-abdallah_fundamental_2010}. The maximum $\beta_{res}$ that satisfies Eq. \eqref{Seq:beta_xi1} occurs near the surface polariton resonance frequency, $\Omega$, such that $\real{{2}/{\chi (\Omega)}}=-1$ or $\displaystyle \real{r_p}=0$.
At $\omega=\Omega$, Eq. \eqref{Seq:beta_xi1} yields 
\begin{equation}
      \betamax\approx\frac{1}{2d}\ln \imag{r_p}^2.
\end{equation} 
The logarithmic dependence of $\beta_\mathrm{max}$ from the imaginary part of the Fresnel coefficient showcases the role of material loss in NFRHT \cite{ben-abdallah_fundamental_2010}.

The emission spectrum $\Phi$ is therefore given by
    \begin{multline}\label{Seq:Phi}
        \Phi\bomega =\frac{1}{4\pi^2}\int_0^\infty \beta[\xi_p (\omega,\beta)+\xi_s (\omega,\beta)]d\beta\\
        \simeq \frac{1}{4\pi^2}\int_{k_0 d}^\infty \beta\xi_p (\omega,\beta)d\beta\simeq\frac{1}{4\pi^2}\int_0^\infty \beta\xi_p (\omega,\beta)d\beta \\
         =\frac{1}{4\pi^2d^2}\int_0^\infty x\xi_p(\omega,x)dx,\quad
    \end{multline}
where we have assumed $k_0 d\ll 1$ and we have made the substitution $\beta d \to x$. By plugging the expression of $\xi_p$ given in Eq. \eqref{Seq:xi_p} in Eq. \eqref{Seq:Phi}, and by making the substitution $e^{-2x}\to y$, we obtain:
\begin{multline}\label{Seq:Phi_demon}
\displaystyle  \Phi =- \frac{\imag{r_p}^2}{4\pi^2d^2} \int_0^{1} \frac{\log y}{|1-r_p^2y|^2}dy\\
 =-\frac{1}{8\pi^2d^2}\frac{\imag{r_p}}{\real{r_p}}\text{Im}\left\{\int_0^{r_p^2} \frac{\log y}{1-y}dy\right.\\
\left.
-\log {r_p^2}\int_0^{r_p^2} \frac{1}{1-y}dy,
\right\}
%  &=\frac{\imag{r_p}}{8\pi^2d^2\real{r_p}}\text{Im}\left\{ \Li{r_p^2} +2 \log{r_p^2} \log{(1-r_p^2)} \right\}.
\end{multline}
\begin{figure}
    \centering
    \includegraphics[width=0.4\columnwidth]{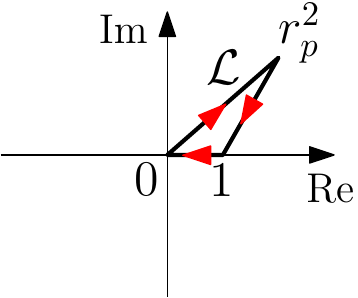}
    \caption{Contour $\mathcal{L}$ in the complex plane for the identity in Eq. \eqref{Seq:int_1}. It is a triangle with vertices $0,\,1,\,r_p^2$, where $r_p$ is the Fresnel coefficient for $p$-polarization.}
    \label{fig:contour1}
\end{figure}

We now solve the last two integrals in Eq. \eqref{Seq:Phi_demon}. 
We note that the function $\frac{\log y}{1-y}$ is analytical  inside the contour $\mathcal{L}$ in the complex plane depicted in Fig. \ref{fig:contour1}. Thus, by applying \emph{Cauchy integral's theorem} \cite{walsh_cauchy-goursat_1933}, $\displaystyle \oint_\mathcal{L}\frac{\log y}{1-y}dy=0$, hence the first integral is solved as:
\begin{multline}\label{Seq:int_1}
    \int_0^{r_p^2} \frac{\log y}{1-y}dy = \int_1^{r_p^2} \frac{\log y}{1-y}dy+\int_0^{1} \frac{\log y}{1-y}dy\\
    =\Li{1-r_p^2} - \frac{\pi^2}{6},
\end{multline}
where $\text{Li}_2$ is the dilogarithm or Spence's function, defined as \cite{lewin_dilogarithms_1958} 
\begin{equation}\label{Seq:dilog}
 \text{Li}_2(z) = -\int_0^z \frac{\log{1-u}}{u}du,\quad\forall z\in\mathcal{C}.
 \end{equation}

The second integral is simply $\displaystyle\int_0^{r_p^2} \frac{\log {r_p^2}}{1-y}dy=-\log{r_p^2}\log{(1-r_p^2)}$. Therefore, by plugging this result and the one of Eq. \eqref{Seq:int_1}, in Eq. \eqref{Seq:Phi_demon}, we can express the thermal emission spectrum, $\Phi$, as:
\begin{multline}\label{Seq:Phi_def}
   \displaystyle  \Phi = \frac{1}{8\pi^2d^2}\frac{\imag{r_p}}{\real{r_p}}\text{Im}\left\{
   -\Li{1-r_p^2} \right.\\
   \left.+\frac{\pi^2}{6}-\log{r_p^2}\log{(1-r_p^2)}
   \right\}\\
   =\frac{1}{8\pi^2d^2}\frac{\imag{r_p}}{\real{r_p}}\text{Im}\left\{\Li{r_p^2}\right\},
\end{multline}
where we have used the dilogarithm identity $\Li{z}+\Li{1-z}=\frac{\pi^2}{6}-\log z\log{(1-z)},\,\forall z\in\mathcal{C}\setminus\{1\}$ \cite{zagier_dilogarithm_2007}. Eq. \eqref{Seq:Phi_def} coincides with Eq. \eqref{eq:Phi_analytical} in the main text, and applies to \textit{any} material dispersion. The crucial difference with \cite{rousseau_asymptotic_2012} is having isolated the term $\real{r_p}$ in the  dilogarithm prefactor's denominator, which is central in deriving the simplified final expression for the radiative thermal conductance in Eq. \eqref{eq:HTC_analytical}.

In the case of the NFRHT between different polaritonic materials with permittivities $\varepsilon_1,\,\varepsilon_2$, the emission spectrum, $\Phi$, can be derived as in Eq. \eqref{Seq:Phi_demon}, and has the following expression:
\begin{equation}\label{Seq:Phi_dissimilar}
    \Phi= \frac{1}{4\pi^2d^2}\left[\frac{\real{r_{p,1}}}{\imag{r_{p,1}}}+\frac{\real{r_{p,2}}}{\imag{r_{p,2}}}\right]^{-1}\text{Im}\left\{\Li{r_{p,1} r_{p,2}}\right\},
\end{equation}
where $r_{p,1},r_{p,2}$ are the Fresnel coefficient at the interfaces with the media of permittivities $\varepsilon_1,\,\varepsilon_2$, respectively.
Eqs. \eqref{Seq:Phi_def} and \eqref{Seq:Phi_dissimilar} agree with \cite{rousseau_asymptotic_2012}.
% \subsection{Low-Loss Limit}

 In the low-loss limit, the polariton resonance frequency $\Omega$ can be found as the real solution of $\real{r_p({\Omega})}=0$ (see Eq. \eqref{eq:resonance_freq}). In this limit, the emission spectrum is maximum at $\omega=\Omega$. As $\displaystyle \lim_{r_p\to 0} \Phi$, we find that \eqref{Seq:Phi_def} simplifies to:
\begin{multline}\label{Seq:MillerBound}
    \Phi_\text{max}=\Phi(\Omega)= \frac{1}{4\pi^2d^2}\ln{\left[1+\imag{r_p(\Omega)}^2\right]}\\
    \approx\frac{1}{4\pi^2d^2}\ln{\left[\left.\frac{|\chi|^4}{4\imag{\chi}^2}\right|_\Omega\right]},
\end{multline}
where the identity $\displaystyle\imag{r_p(\Omega)}=\left.\frac{|\chi|^2}{2\imag{\chi}}\right|_{\omega=\Omega}$ was used. Eq. \ref{Seq:MillerBound} agrees exactly with the result by Miller \textit{et al}. in \cite{miller_shape-independent_2015} (Eq. (10)), derived for  planar configurations. By contrast, in the high-loss limit, Eq. \eqref{eq:resonance_freq} may have no real solutions, $\Omega$, and the maximum of $\Phi$ needs to be calculated by maximizing the right hand side in Eq. \eqref{Seq:Phi_def}. The range of frequencies for which Eq. \eqref{Seq:MillerBound} is valid and the threshold between low-loss and high-loss regimes is discussed in the following section.

\section{Exact derivation of the radiative thermal conductance}\label{appendixB}
The radiative thermal conductance for two closely spaced semi-infinite layers, given in Eq. \eqref{eq:HTC_def}  can be rewritten as:
\begin{equation}\label{seq:HTC_def}
    h=\int_0^{\infty}k_B \Pi\left(\frac{\omega}{T}\right) \Phi\bomega d\omega,
\end{equation}
where $\Pi$ is given in Eq. \eqref{eq:PiT}, and $\Phi\bomega $ is the emission spectrum, whose closed form expression has been derived in the previous section and is given in Eq. \eqref{Seq:Phi_def}.

We now particularize this derivation to plasmonic and polar media, whose dispersion relations are given in Eqs. \eqref{eq:eps_plasm} and \eqref{eq:eps_pol}, respectively.

We assume that the function $\Pi\left(\frac{\omega}{T}\right)$ is slowly varying with respect to the emission spectrum $\Phi(\omega)$, which peaks at the polariton resonance frequency $\bar\Omega$. Therefore, we make the first step toward the analytical integration of Eq. \eqref{seq:HTC_def} by sampling $\Pi$ at the frequency $\bar\Omega$, i.e.:
$\displaystyle
    h=k_B \Pi\left(\frac{\bar\Omega}{T}\right)\int_0^{\infty} \Phi\bomega d\omega.
$

We now carry out the frequency integration of the emission spectrum.
By using its expression in Eq. \eqref{Seq:Phi_def}, we have
\begin{equation}\label{Seq:h_step1}
8\pi^2d^2\int_0^\infty\Phi(\omega)d\omega=\int_0^\infty\frac{\imag{r_p\bomega}}{\real{r_p\bomega}}\text{Im}\left\{\Li{r_p^2\bomega}\right\}d\omega.
\end{equation}
Since $r_p(\omega)$, inherits the hermiticity (or PT-symmetry) from the permittivity function $\varepsilon(\omega)$, i.e. $r_p^*(\omega)=-r_p(-\omega)$ ($^*$ is the complex-conjugate operator), the integrand in Eq. \eqref{Seq:h_step1} is an even function of $\omega$. Therefore, we can extend the integration to include the negative frequency axis, namely
\begin{equation}\label{Seq:h_step2}
8\pi^2d^2    \int_0^\infty\Phi(\omega)d\omega=\text{Im}\left\{\int_{-\infty}^{+\infty}f(\omega)d\omega\right\},
\end{equation}
where the complex valued function $f(z)$ is
\begin{equation}\label{Seq:f}
f(z)=    \frac{\imag{r_p}(z)}{2\real{r_p}(z)}\Li{r_p^2(z)}.
\end{equation}

\begin{figure}[t]
    \centering
    \includegraphics[width=0.9\columnwidth]{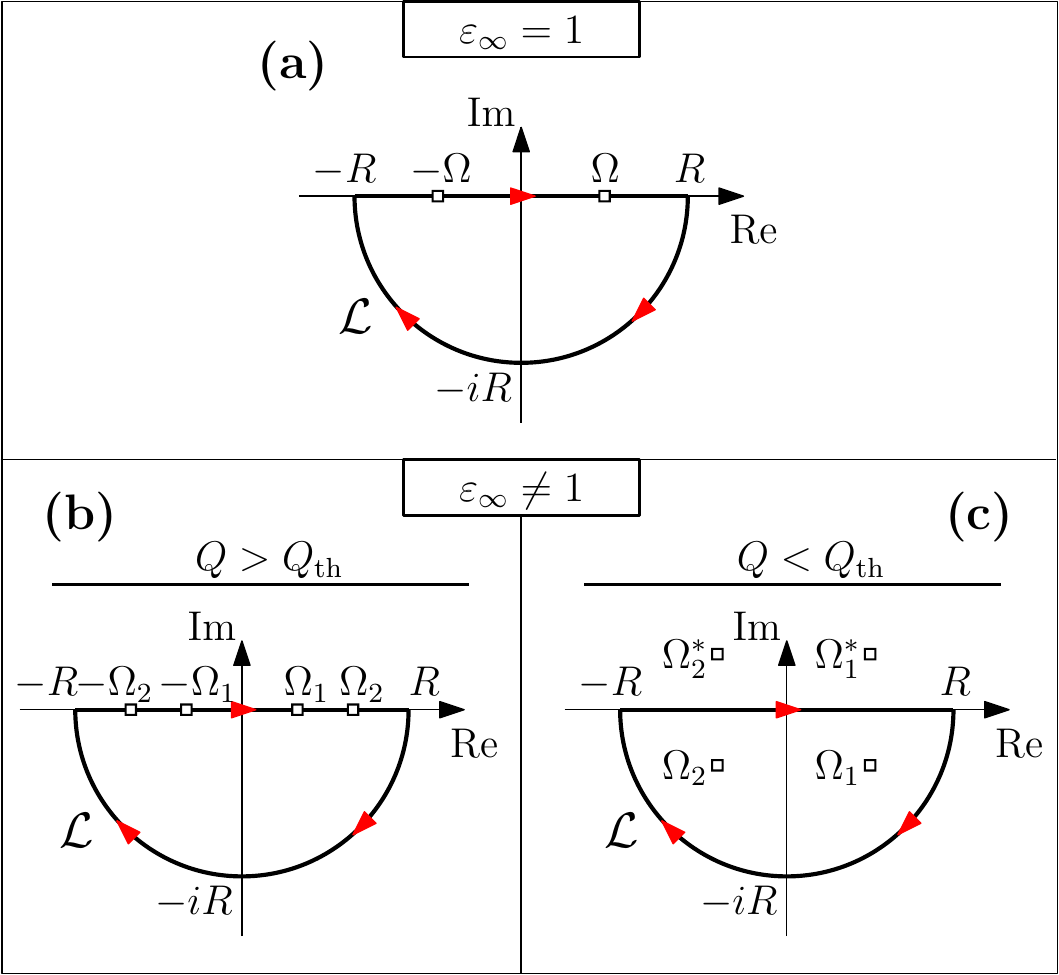}
    \caption{Contour $\mathcal{L}$ in the complex plane for the complex contour integration in Eq. \eqref{Seq:h_step3}, composed by the real axis and a semicircular contour of radius $R$, which will tend to infinity in order to cover the lower half of the complex plane. The poles of the function $f(z)$, defined in Eq. \eqref{Seq:f}, are shown for the cases $\varepsilon_\infty=1$ {\bf(a)}, $(\varepsilon_\infty\neq1,Q>\Qth)$ {\bf(b)} and $(\varepsilon_\infty\neq1,Q<\Qth)$ {\bf(c)}.}
    \label{fig:contour2}
\end{figure}
We now tackle the complex integration of $f\bomega$ in Eq. \eqref{Seq:h_step2} by means of contour integration in the complex plane, following a similar strategy to the one empolyed in \cite{rousseau_asymptotic_2012}. Specifically, we intend to perform the integration on the closed contour $\mathcal{L}$ shown in Fig. \ref{fig:contour2}, composed by the real axis and a semicircular contour of positive radius $R$ lying the lower half-plane, in the limit $R\to\infty$. 
Since the integrand is vanishing on the semicircular contour in the limit $\displaystyle\lim_{R\to\infty}$, from Jordan's lemma \cite{carrier_functions_2005}, the integral on this contour is also zero. Hence, we can rewrite the integrated emission spectrum as:
\begin{equation}\label{Seq:h_step3}
    8\pi^2d^2    \int_0^\infty\Phi(\omega)d\omega    =\text{Im}\left\{\oint_\mathcal{L}f(z)dz\right\}.
\end{equation}
The function $f(z)$ is the analytical extension of the real-valued function $f\bomega$ to the complex plane. It must be noted that  $\imag{r_p}(z),\,\real{r_p}(z)$  are no longer constrained to be real valued functions, and therefore the notation $\imag{\cdot},\,\real{\cdot}$ no longer refers to the real and imaginary part operators. Nevertheless, we keep using the same notation in the following calculations for the sake of simplicity, while accounting that $\imag{r_p}(z),\,\real{r_p}(z)$ are functions derived for real variable, and extended to the complex plane. For instance, the step $\imag{r_p\bomega}\to\imag{r_p}(z)$ for plasmonic dispersions is the following:
\begin{multline}\label{Seq:imagrp}
    \imag{r_p\bomega}\\
    =\frac{2\varepsilon_\infty \omega\omega_p^2 \gamma}{[(1+\varepsilon_\infty)\omega^2-\varepsilon_\infty \omega_p^2]^2+(1+\varepsilon_\infty )^2\omega^2\gamma^2}\\
    \to
    \frac{2\varepsilon_\infty z\omega_p^2 \gamma}{[(1+\varepsilon_\infty)z^2-\varepsilon_\infty \omega_p^2]^2+(1+\varepsilon_\infty )^2z^2\gamma^2}\\
    =\imag{r_p}(z),\qquad z\in\mathcal{C}.
\end{multline}

We now carry the complex integration using the Cauchy's residue theorem \cite{carrier_functions_2005}. The first step is identifying the poles of the integrand $f(z)$. By inspecting the expression of $f(z)$ in Eq. \eqref{Seq:f}, it is clear that the poles are exactly the frequencies $\bar\Omega$ solving the resonance condition in Eq. \eqref{eq:resonance_freq}. By solving Eq. \eqref{eq:resonance_freq}, using the expression of the plasmonic and polar permittivity given in Eq. \eqref{eq:plasm_pol_dispers}, we have:
\begin{equation}\label{Seq:OmegaGeneral_plasm}
    \bar\Omega_{\substack{1 \\ 2}}=\Omega,\qquad\varepsilon_\infty=1, 
\end{equation}   
while
\begin{equation}\label{Seq:OmegaGeneral_pol}
    \bar\Omega_{\substack{1 \\ 2}}=
     \displaystyle  \Omega\sqrt{F-\frac{1}{2Q^2}\mp \frac{1}{2} \sqrt{\left(\frac{1}{\Qth^2}-\frac{1}{Q^2}\right)\left(\frac{1}{Q_2^2}-\frac{1}{Q^2}\right)}}
\end{equation}
for $\varepsilon_\infty \neq 1$. Here,  $\Omega$ is the polariton resonance frequency in the absence of optical losses, given in Eq. \eqref{eq:omega_res_gen}, and $Q$ is the quality factor of the polaritonic material resonance, defined in Eq. \eqref{eq:Q}.
The parameter F for plasmonic and polar cases can be written as:
\begin{subnumcases}{F=\label{Seq:F}}
         \displaystyle   \Fplasm= 
        %  \frac{\varepsilon_\infty}{\varepsilon_\infty-1}
        1+\frac{1}{2(\Bplasm-1)}
         \label{Seq:F_plasm}\\
         \displaystyle \Fpolar =1+\frac{\varepsilon_\infty+1}{2\Bpolar(\varepsilon_\infty-1)}  \label{Seq:F_pol}. 
\end{subnumcases}
The parameter $B$ is the material residue function, independent of the material losses (independent of $\gamma$), given in Eq. \eqref{eq:Bplasm_pol}. 
Finally, the parameters $\Qth$ and $Q_2$ have the following expressions:
\begin{align}
    \Qth &= \frac{1}{\sqrt{2\left(F-\sqrt{2F-1}\right)}}\label{Seq:Qth}\\
     Q_2 &= \frac{1}{\sqrt{2\left(F+\sqrt{2F-1}\right)}}\label{Seq:Q2},
\end{align}
where $F$ for plasmonic and polar dispersions is given in Eq. \eqref{Seq:F}.

From Eq. \eqref{Seq:OmegaGeneral_pol}, in the case $\varepsilon_\infty \neq 1$, it can be shown that $\bar\Omega_{1,2}$ are real numbers only if $Q<Q_2$ and $Q>\Qth$. It can be proven that $Q_2<1$, and hence we can focus only on the cases $Q<\Qth$ and $Q>\Qth$. Therefore, $\Qth$ represents a threshold for the material quality factor below which the poles of $f(z)$ become complex and move away from the real axis, as shown in Figs. \ref{fig:contour2}b-c. 

We can now apply the residue theorem to carry out the complex integration in Eq. \eqref{Seq:h_step3}. It is important to notice that, since for $\varepsilon_\infty\neq 1$ and $Q>\Qth$ the poles of $f(z)$ are on the integration contour, we have to add a $\frac{1}{2}$ factor to the standard residue theorem formula, for which the poles are in the interior of the integration contour $\mathcal{L}$ \cite{carrier_functions_2005}. On the other hand, the frequency $\bar\Omega$ in the case $\varepsilon_\infty=1$ does not depend on the $Q$ factor, and the poles are always real.\\

{\bf Case $\varepsilon_\infty=1$. }\\
% \subsection{Case $\varepsilon_\infty=1$}
Via algebraic manipulation, it can be shown that the function $f(z)$ for both plasmonic and polar dispersions in the case $\varepsilon_\infty=1$ has the following expression:
\begin{equation}
    f(z) = \frac{\Omega}{2Q}\frac{z}{z^2-\Omega^2}\Li{r_p^2(z)}.
\end{equation}
Therefore, $f(z)$ has two first-order real poles, i.e. $\{+\Omega,-\Omega\}$, and the integration of Eq. \eqref{Seq:h_step3} can be carried out through the residue theorem as follows:
\begin{multline}\label{Seq:h_step4_einf1}
 \text{Im}\left\{\oint_\mathcal{L}f(z)dz\right\} \\
 = 
 -\text{Im}\left\{ i\pi \left[ \res (f,+\Omega)+\res (f,-\Omega)\right]\right\}\\
 =-\text{Re}\left\{2\pi  \res (f,+\Omega) \right\},
\end{multline}
where $\res (f,w)$ is the residue of $f$ at $w$. Here, we have used the parity of $f$, and the minus sign comes from having chosen a clockwise (negative) orientation of the contour $\mathcal{L}$, shown in Fig. \ref{fig:contour2}a.

By taking the limit $\displaystyle \lim_{z\to\Omega}(z-\Omega)f(z)$, we calculate the residue $\res (f,+\Omega)$, which has the following expression:
\begin{multline}\label{Seq:residue_einf1}
    \res (f,+\Omega)= \frac{\Omega}{4Q}\Li{-\imag{r_p(\Omega)}^2}\\
    =\frac{\Omega}{4Q}\Li{-\left(\frac{Q}{B}\right)^2}.
\end{multline}
By plugging this result in Eq. \eqref{Seq:h_step4_einf1}, we can finally write the expression for the radiative thermal conductance as in Eq. \eqref{eq:HTC_analytical}. It must be noted that all the redundant scaling factors, e.g. $B$ at the denominator of $\hmax$ and $\Psi$, have been introduced such that $\Psi$ would be bounded above by 1.

{\bf Case $\varepsilon_\infty\neq 1$. }\\
The function $f(z)$ for both plasmonic and polar dispersions in the
case $\varepsilon_\infty\neq 1$ has the following expression:
\begin{equation}\label{Seq:f_einfnot2}
    f(z) = \frac{(F-1)\Omega^3 z}{(z^2-\bar\Omega_1^2)(z^2-\bar\Omega_2^2)}\frac{\Li{r_p^2(z)}}{Q},
\end{equation}
where $F$ is given in Eq. \eqref{Seq:F}.

For $\varepsilon_\infty\neq 1$, we have different resuts according to the position of $Q$ with respect to $\Qth$. From Eq. \eqref{Seq:OmegaGeneral_pol}, it is clear that if $Q>\Qth$, then the function $f(z)$ has four first-order real poles $\{\pm \bar\Omega_1,\pm \bar\Omega_2\}$, as shown in Fig. \ref{fig:contour2}b; if $Q<\Qth$ then the function $f(z)$ has four first-order complex poles $\{\pm\bar\Omega_1,\pm \bar\Omega_2\}$, with $\bar\Omega_2=-\bar\Omega_1^*$ as shown in Fig. \ref{fig:contour2}c. Thus, if $Q<\Qth$, the only poles contributing to the integral are the two in the interior of the contour $\mathcal{L}$, viz. $\{\bar\Omega_1,\,\bar\Omega_2\}$. Therefore, in both cases $Q>\Qth$ and $Q<\Qth$ we can solve Eq. \eqref{Seq:h_step3} by applying the residue theorem as follows:
\begin{equation}\label{Seq:int_general_last}
    \text{Im}\left\{\oint_\mathcal{L}f(z)dz\right\}=-\text{Re}\left\{2\pi \sum_{j=1}^2 \res (f,\bar\Omega_j) \right\}.
\end{equation}
By making the limit $\displaystyle \lim_{z\to\Omega_j}(z-\bar\Omega_j)f(z)$, we calculate the residues $\res (f,\bar\Omega_1)$, $\res (f,\bar\Omega_2)$  which have the following expression:
\begin{multline}\label{Seq:f_einfnot1}
     \res(f,\bar\Omega_{\substack{1 \\ 2}})  = \mp\frac{(F-1) }{2\sqrt{4(F^2-1)+\frac{1}{Q^4}-4F(2+\frac{1}{Q^2})}}\\   
    \times\frac{\Li{-\imag{r_p}(\bar\Omega_{\substack{1 \\ 2}})^2}}{Q}\qquad
\end{multline}%
where $F$ is given in \eqref{Seq:F}, and the complex function $\imag{r_p}(z)$ is the analytical extension of the imaginary part of the Fresnel coefficient in the complex plane (e.g., see Eq. \eqref{Seq:imagrp} for the plasmonic dispersion). By inserting the residues in Eq. \eqref{Seq:f_einfnot1} into Eq. \eqref{Seq:int_general_last}, and in turn plugging this into Eq. \eqref{Seq:h_step3}, one can finally get the expression for the heat conductance.

\begin{figure}[h!]
    \centering
    \includegraphics[width=\columnwidth]{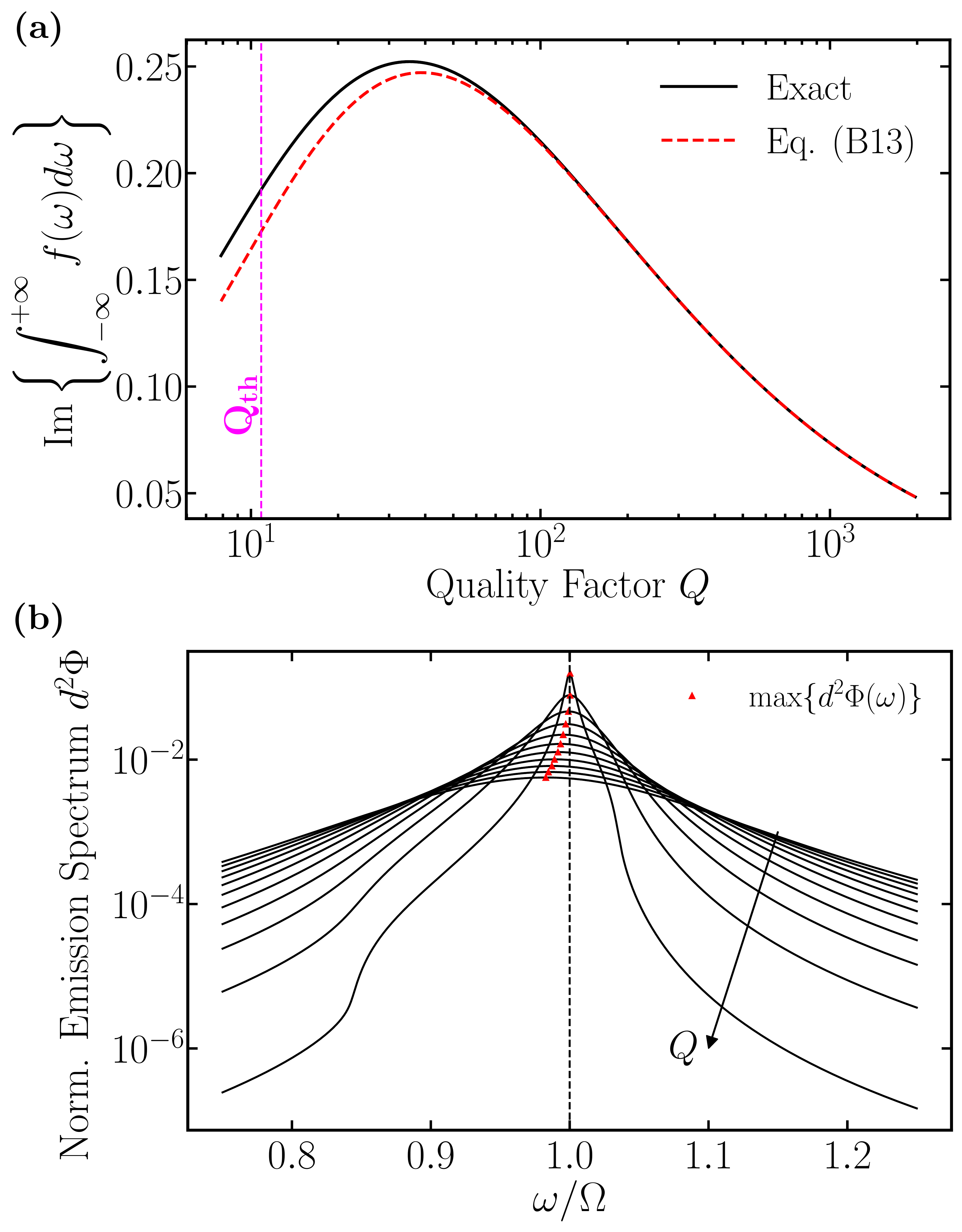}
    \caption{{\bf(a)} Integral in Eq. \eqref{Seq:h_step3}, equivalent to $8\pi^2d^2\int_0^\infty\Phi(\omega)d\omega$, for a polar dispersion with Lorentz parameters $\varepsilon_\infty=4$, $\omega_\text{TO}=1.49\times 10^{14}\rads$, $\omega_\text{LO}=1.83\times 10^{14}\rads$, and resonance frequency $\Omega=1.77\times 10^{14}\rads$. We compare the exact solution (black curve) calculated using Eq. \eqref{Seq:int_general_last}, with the approximated one (red dashed line) calculated using Eq. \eqref{Seq:h_step4_einf1}, and used as final result in the main text.
    We also show with a vertical magenta line the threshold $\Qth=10.85$ between the low- and high-loss regimes, calculated using Eq. \eqref{Seq:Qth}.
    For the same dispersion parameters, in {\bf (b)} we plot the emission spectrum normalized by $d^{-2}$, for increasing values of $Q=\Omega/\gamma$, namely $Q=Q_1/50, Q_1/45,Q_1/40,\dots,Q_1$, being $Q_1\approx 200$. We also mark the peak of each emission spectrum with a red marker: in every case, the resonance occurs at frequencies very close to $\Omega$, calculated for $Q\to\infty$.}
    \label{fig:fig3S}
\end{figure} 
\section{Approximations leading to Eq. \eqref{eq:HTC_analytical}} \label{appendixC}
We now simplify the exact expressions for $h$, derived in the previous section, for the three scenarios, viz. $\{\varepsilon_\infty=1,\forall Q\}$, $\{\varepsilon_\infty\neq1, Q<\Qth\}$ and $\{\varepsilon_\infty\neq1, Q>\Qth\}$, aiming at providing a single expression valid in any regime of $Q$ and $\varepsilon_\infty$.

We make the following approximations: (\textit{i}) we neglect the contribution from the second pole $\bar\Omega_2$; (\textit{ii}) we assume $Q\gg\Qth$. Under these assumptions, we can approximate $\bar\Omega_1\approx\Omega$, with $\Omega$ given in Eq. \eqref{eq:omega_res_gen}. It can be shown that the resulting residue $\res(f,\Omega)$ has the same form as the case $\varepsilon_\infty=1$ in Eq. \eqref{Seq:residue_einf1}, and the heat transfer conductance expression is the same as in Eq. \eqref{eq:HTC_analytical}. 

Even if this expression is derived in the high-$Q$ limit, it represents a good approximation also in the low-$Q$ case, as shown in Fig. \ref{fig:fig3S} (a) for a case of study. Specifically, we show the integral in Eq. \eqref{Seq:h_step3} for a polar dispersion (see the Lorentz parameters in the figure caption) as a function of the quality factor $Q$, and there is good agreement with the exact solution over all the considered $Q$ range.

It must be noted that the polariton resonance frequency $\bar\Omega_1$ given in Eq. (\ref{Seq:OmegaGeneral_plasm},~\ref{Seq:OmegaGeneral_pol}) for both plasmonic and polar dispersions is weakly dependent from the optical loss, i.e. $\gamma$ or the quality factor, $Q$. In Fig. \ref{fig:fig3S} (b) we show this by monitoring the peak position of the emission spectrum, $\Phi\bomega,$ for the same polar dispersion used in panel (a), for decreasing values of the quality factor, $Q$, or equivalently for increasing values of $\gamma$, starting from $Q_1\approx 200$ and reaching $Q=Q_1/50\approx 4$. It is clear that even in the lowest-$Q$ case, the emission spectrum peaks very closely to the resonance frequency $\Omega$ calculated in the limit $\gamma\to0$ or $Q\to \infty$, given in Eq. \eqref{eq:omega_res_gen}. Thus, assuming $\Omega$ as the polariton resonance frequency in any material loss condition, represents a good approximation.

% \section{Comparison with literature \cite{ben-abdallah_fundamental_2010}}
We now compare the near-field radiative thermal conductance derived in this work with the expression for polar dielectrics provided by Ben-Abdallah \etal
in \cite{ben-abdallah_fundamental_2010}. In \cite{ben-abdallah_fundamental_2010}, the authors considered a polar material dispersion with high-frequency dielectric permittivity $\varepsilon_\infty=1$. In their derivation (Eq. (14)), the radiative thermal conductance that they obtained, $h'$, can be written as:
\begin{equation}\label{Seq:benAbd_HTC}
    h' = \hmax'\,\Psi'(Q)\,\PiT,
\end{equation}
where $\PiT$ is the same as in Eq. \eqref{eq:HTC_analytical}, given in Eq. \eqref{eq:PiT}, and the functions $\hmax'$ and $\Psi'$ are given by:
\begin{equation}\label{Seq:hmax_Psi_prime}
    \hmax' = \frac{0.12 k_B}{d^2}\Omega, \quad\Psi'(Q)=\frac{\log Q}{0.37\,Q}.
\end{equation}
In both our result (Eq. \eqref{eq:HTC_analytical}) and the result from \cite{ben-abdallah_fundamental_2010} (Eq. \eqref{Seq:benAbd_HTC}), since $(\Psi,\Psi')$ and $\Pi$ are functions bounded above by 1, $\hmax$ and $\hmax'$ represent the maximum heat transfer conductance achievable in this configuration. As we shown in the previous sections, the material residue, $B$, is greater than unity, i.e. $B>1$, therefore we can compare $\hmax$ and $\hmax'$ as follows:
${\hmax'}/{\hmax}=4.44\,B>4.4.$
Therefore our analytical estimation for the maximum radiative thermal conductance is at least 4.4 times smaller than the one predicted in \cite{ben-abdallah_fundamental_2010} under the assumption of blackbody-like thermal emission in the near-field ($\xi=1$).

We now compare the functions $(\Psi,\Psi')$, taking into account the dependence of heat transfer from optical loss. The function $\Psi'$ depends only on the quality factor $Q$ of the polariton resonance, and neglects the dependence from the other features of a material's dispersion, such as the size of the Reststrahlen band ($\omega_\text{LO}-\omega_\text{TO}$) and its position. In our work, these are included in the material residue, $B$. According to the definition of $\Psi'(Q)$, in Eq. \eqref{Seq:hmax_Psi_prime}, this function is maximized at the optimal quality factor  $\Qopt^*=e$, being $e\approx2.72$ the Napier's constant, valid for any polar dielectric with a Lorentz dispersion relation (Eq. \eqref{eq:plasm_pol_dispers}), with $\varepsilon_\infty=1$. Conversely, from our derivation, the function $\Psi$ depends on the dispersion's characteristics through the factor $B$, and the optimal quality factor $Q$ is given by $\Qopt=4.5\,B$. 
\begin{figure}[t]
    \centering
    \includegraphics[width=\columnwidth]{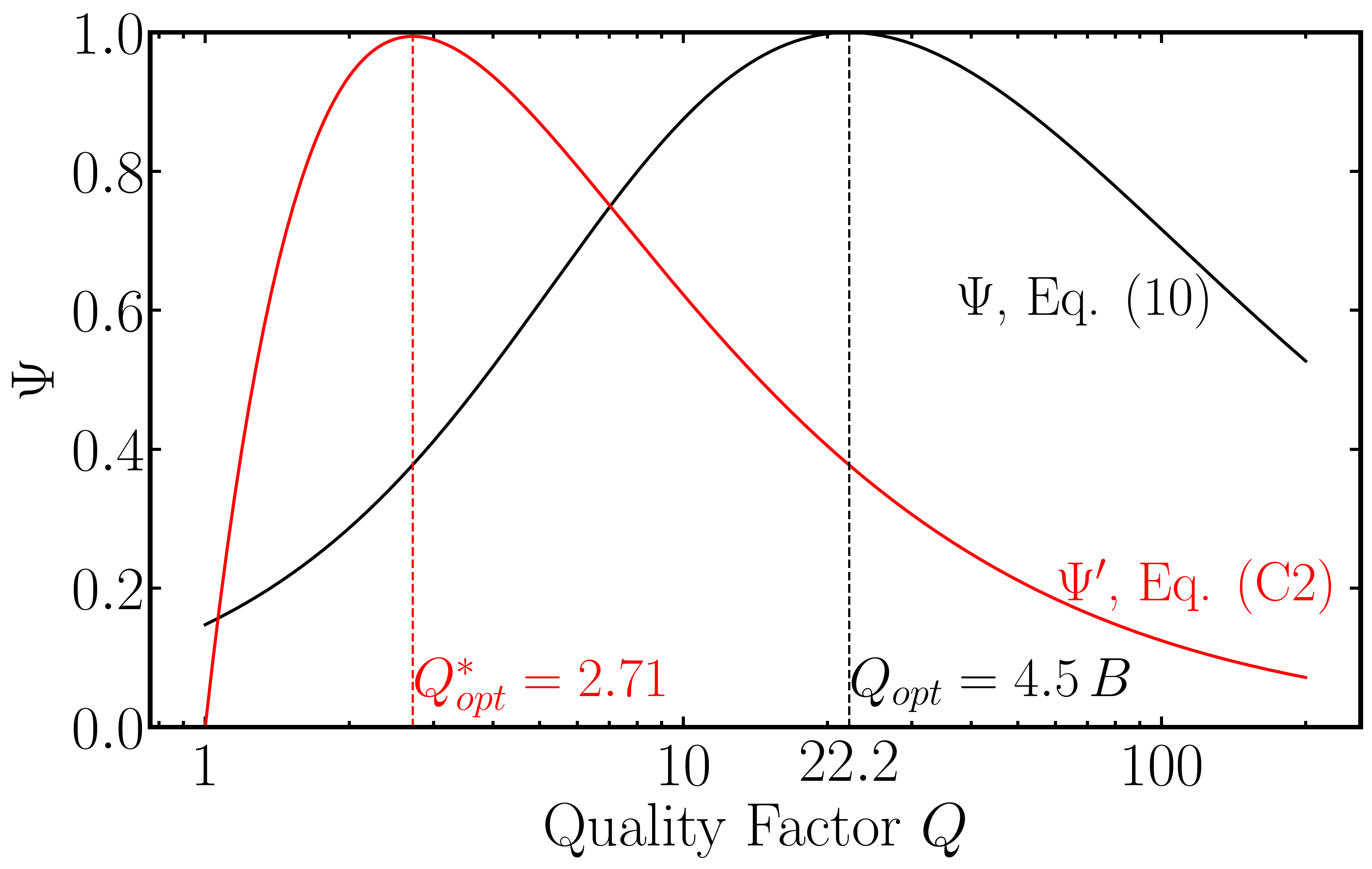}
    \caption{Functions $\Psi$ (black curve) and $\Psi'$ (red curve), accounting for the optical loss for the NFRHT in a polar bulk system with small vacuum gap, calculated using our result in Eq. \eqref{eq:Psi} and the expression from Ben-Abdallah \etal \cite{ben-abdallah_fundamental_2010} given in Eq. \eqref{Seq:hmax_Psi_prime}, respectively,  as a function of the polariton resonance quality factor $Q$. We considered a Lorentz dispersion with parameters $\varepsilon_\infty=1$, $\omega_\text{TO}=1.49\times10^{14}\rads$, $\omega_\text{LO}=1.83\times10^{14}\rads$, for which $B=4.93$, calculated using Eq. \eqref{eq:Bpol}. The different optimal quality factors  at which the curves peak are also marked with a dashed line.}
    \label{fig:Psi_PsiP}
\end{figure}

In Fig. \ref{fig:Psi_PsiP}, we compare the two functions for a Lorentz dispersion with $\varepsilon_\infty=1$, $\omega_\text{TO}=1.49\times10^{14}\rads$, $\omega_\text{LO}=1.83\times10^{14}\rads$, for which $B=4.93$, calculated using Eq. \eqref{eq:Bpol}. The optimal quality factor predicted by maximizing $\Psi$ in Eq. \eqref{eq:Psi} is $\Qopt=4.5\,B=22.2$, about an order of magnitude greater than the optimal $Q$ obtained maximizing $\Psi'$ in Eq. \eqref{Seq:hmax_Psi_prime}.
% \bibliography{references_comb3}
\input{references_final.bbl}

\end{document}